\documentclass[twocolumn,aip,nofootinbib,jcp,reprint,amsmath,amssymb]{revtex4-1}

\usepackage{natbib,hyperref}
\usepackage{amsmath}
\usepackage{graphicx}
\usepackage{hyperref} 
\usepackage{dcolumn}
\usepackage{bm}
\usepackage{epsfig,color,xspace,multirow,xr,bbold}
\usepackage[all]{xy}
\usepackage{setspace}
\usepackage{url}
\usepackage{soul}
\usepackage[colorinlistoftodos]{todonotes}
\usepackage{longtable}
\usepackage{tabularx}
\usepackage{adjustbox}
\usepackage{threeparttable} 
\usepackage{natbib,hyperref}
\usepackage{float}
\usepackage{placeins}
\usepackage[utf8]{inputenc}

\newcommand{\cs}{$C_{\rm s}$\,}
\newcommand{\ctwov}{$C_{\rm 2v}$\,}
\newcommand{\cthreeh}{$C_{\rm 3h}$\,}
\newcommand{\cthreev}{$C_{\rm 3v}$\,}
\newcommand{\dthreeh}{$D_{\rm 3h}$\,}
\newcommand{\ie}{{\it i.e.,}\,}

\newcommand{\etal}{{\it et al.\,}}

\usepackage[normalem]{ulem}
\usepackage{xcolor}
\newcommand{\RRef}[1]{Ref.~\onlinecite{#1}}

\newcommand\notsotiny{\@setfontsize\notsotiny\@vipt\@viipt}
\usepackage{soul} 
\begin{document}

\title{
Influence of Pseudo-Jahn--Teller Activity on
the Singlet-Triplet Gap of Azaphenalenes
}
\date{\today}

\author{Atreyee Majumdar}
\author{Komal Jindal}
\author{Surajit Das}
\author{Raghunathan Ramakrishnan}

\email{ramakrishnan@tifrh.res.in}
\affiliation{$^1$Tata Institute of Fundamental Research, Hyderabad 500046, India}


\begin{abstract}
\noindent
We analyze the possibility of symmetry-lowering induced by pseudo-Jahn--Teller interactions in six previously studied azaphenalenes that are known to have their first excited singlet state (S$_1$) lower in energy than 
the triplet state (T$_1$). 
The primary aim of this study is to explore whether Hund's rule violation is observed in these molecules when their structures are distorted 
from  \ctwov or \dthreeh point group symmetries
by vibronic coupling.
Along two interatomic distances connecting these point groups to their subgroups \cs or \cthreeh, we 
relaxed the other internal degrees of freedom and 
calculated two-dimensional potential energy subsurfaces. 
The many-body perturbation theory (MP2) suggests that the high-symmetry structures are the energy minima for all six systems. However, single-point energy calculations using the coupled-cluster method (CCSD(T)) indicate symmetry lowering in four cases.
The singlet-triplet energy gap plotted on the potential energy surface also shows variations when deviating from high-symmetry structures. A full geometry optimization at the CCSD(T) level with the cc-pVTZ basis set reveals that the \dthreeh structure of cyclazine (1AP) is a saddle point, connecting two equivalent minima of \cthreeh symmetry undergoing rapid automerization. The combined effects of symmetry lowering and high-level corrections result in a nearly zero singlet-triplet gap for the \cthreeh structure of cyclazine.
Azaphenalenes containing nitrogen atoms at electron-deficient sites---2AP, 3AP, and 4AP---exhibit more pronounced in-plane structural distortion; the effect is captured by the long-range exchange-interaction corrected DFT method, $\omega$B97XD. 
Excited state calculations of these systems indicate that in their low-symmetry energy minima, T$_1$ is indeed lower in energy than S$_1$, upholding the validity of Hund's rule. 
Jahn--Teller analysis predicts the symmetries of the in-plane distortion vibrational modes as $A_2^\prime$:\dthreeh$\rightarrow$\cthreeh  or $B_2$: \ctwov$\rightarrow$\cs agreeing with the vibrational frequencies of the saddle-points.
\end{abstract}

\maketitle

\section{Introduction\label{sec_intro}}

\begin{figure}[H]
\centering
\includegraphics[width=\linewidth]{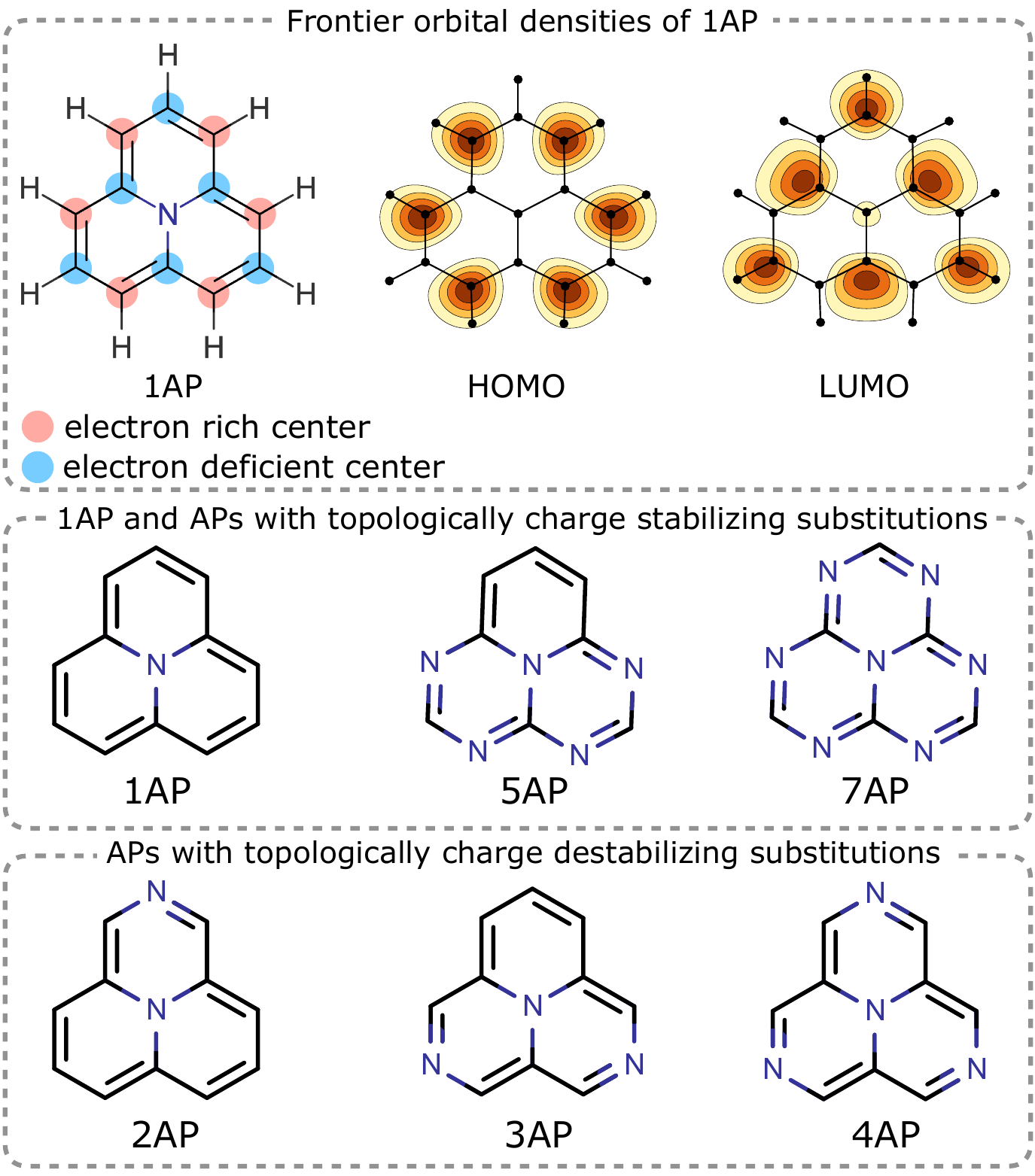}
\caption{
Representation of APs explored in 
the present study, and density contours of the frontier MOs of 1AP. 
}
\label{fig_systems}
\end{figure}

Azaphenalene (1AP in FIG.~\ref{fig_systems}), also 
known as mono-azaphenalene, cycl[3.3.3]azine, or cyclazine, 
and its poly-aza analogues, wherein nitrogen (N) atoms replace the peripheral CH groups, display a
potential to violate the Hund's rule\cite{kutzelnigg1996hund}---their first excited singlet state, S$_1$, is suggested to be lower in energy 
than the triplet counterpart, T$_1$\cite{de2019inverted,ehrmaier2019singlet,pollice2021organic,aizawa2022delayed,li2022organic,ricci2022establishing,ghosh2022origin,tuvckova2022origin,bedogni2023shining,bedogni2023shining,won2023inverted,drwal2023role,loos2023heptazine,pollice2024rational,garner2024enhanced,valverde2024computational,pollice2024rational}. 
Three-fold symmetric 1AP\cite{cunningham1969heterocyclic} and heptazine (7AP)\cite{hosmane1982synthesis,shahbaz1984tri} were among the first APs synthesized, followed by the two-fold symmetric, pentazine (5AP), along with a few other asymmetric APs\cite{rossman1985synthesis}; see FIG.~\ref{fig_systems} for the structures of 5AP and 7AP. Gimarc\cite{gimarc1983topological} ascribed the stability of 7AP to the substitution of the CH moiety by N at the sites of 1AP with significant densities of the highest occupied molecular orbital (HOMO), as shown in FIG.~\ref{fig_systems}. 
This topological charge stabilization effect\cite{gimarc1983topological}  
rationalized the synthesis of another three-fold symmetric AP\cite{leupin19861}. 
A similar effect diminishes anti-aromaticity through double-bond localizations in substituted indacene\cite{heilbronner1987influence} and pentalene\cite{terence2023symmetry,omar2023identification,meiszter2024revisiting} that, along with APs, are among the too few
molecular cores identified so far to exhibit negative STGs. A search for other examples in the 
chemical space of small organic molecules showed no evidence of STG inversion\cite{majumdar2024resilience}. 

Leupin \etal were the first to suggest the likelihood of a negative S$_1-$T$_1$ gap (STG) in 7AP\cite{leupin1980low}.
Its photochemistry exhibits delayed fluorescence, characteristic of a 
small, negative STG\cite{ehrmaier2019singlet}. Further, fluorescence measurements 
suggested an STG of $-$0.011 eV for a substituted 7AP\cite{aizawa2022delayed}. 
Finite-temperature Wigner phase space sampling of two experimentally studied derivatives of 7AP has revealed the out-of-plane distortion of the equilibrium geometry of the S$_1$ state to play a role in the downhill reverse-intersystem-crossing (rISC) mechanism\cite{karak2024reverse}.
More recently, an absorption spectrum confirmed the
STG of 5AP in argon matrix to be $-0.047\pm0.007$ eV\cite{wilson2024spectroscopic}
aligning with independent measurements on dialkylamine-substituted 5AP\cite{kusakabe2024inverted}. 
Inverted STG promises the scope for designing the next generation organic light emitting diodes (OLEDs), with efficiencies not limited by spin statistics\cite{aizawa2022delayed,li2022organic,won2023inverted}.

In this study, we investigate structural preferences exhibited by APs shown in FIG.~\ref{fig_systems} and explore how symmetry lowering, driven by second-order Jahn--Teller (SOJT) interactions, impacts their STGs.
We first focus on 1AP, 5AP, and 7AP for which
Loos \etal\cite{loos2023heptazine} performed detailed calculations of STGs and provided the theoretical best estimates (TBEs) as 
$-0.131$ eV, $-0.119$ eV, and $-0.219$ eV, respectively.
We also examine the three APs---2AP, 3AP, and 4AP---containing N at electron-deficient sites of 1AP (see FIG.~\ref{fig_systems}). 
Though these molecules lack experimental measurements, they are of theoretical interest. 
Loos \etal\cite{loos2023heptazine} also reported TBEs of their STGs as $-0.071$ eV (2AP), $-0.042$ eV (3AP), and $-0.029$ eV (4AP). 
In the calculations of TBEs\cite{loos2023heptazine}, geometries were determined using the accurate method, CCSD(T), by enforcing the maximum symmetry point group for 
each composition: 
\dthreeh for 1AP, 4AP, and 7AP;  
\ctwov for 2AP, 5AP, and 3AP. For this purpose, MP2 minimum energy structures verified through frequency analysis were selected as initial geometries for subsequent geometry refinement at the CCSD(T) level of theory. 
Other studies have also applied geometries 
determined with MP2\cite{de2019inverted,tuvckova2022origin} or using DFT methods such as B3LYP\cite{ghosh2022origin} and $\omega$B97X-D\cite{garner2024enhanced}.

Elaborate analysis using wavefunction methods and large basis sets have scrutinized the validity of Hund's rule violation by APs\cite{dreuw2023inverted}. 
Other works have shown the 
limitations of time-dependent DFT (TD-DFT) with functionals lacking a
correlation term at the  MP2 level to predict negative STGs of APs\cite{ghosh2022origin,kondo2022singlet,tuvckova2022origin,sancho2022violation,loos2023heptazine}.
The nature of the potential energy surface (PES) of APs and its 
impact on their STGs has received less attention. 
For 1AP,  early calculations\cite{leupin1980low} suggested a \cthreeh structure with alternating bond lengths, contrasting the \dthreeh structure typical of an aromatic system. 
However, calculations using DFT\cite{de2019inverted} and MP2\cite{ehrmaier2019singlet} methods predicted a symmetric structure, 
which has since been widely adopted. 
It is interesting to note that MP2 predicts a planar, \dthreeh structure for phosphaphenalene containing a phosphorous atom in the triangulene core of 1AP instead of N, while the B97-3C-DFT method indicates a distortion of the geometry towards the 
pyramidal, \cthreev structure\cite{ricci2022establishing}.
Although more accurate calculations of 1AP have been performed using the CCSD(T) method\cite{loos2023heptazine}, vibrational analysis at this level remains challenging when using a triple-zeta quality basis set. 
Additionally, determining an initial geometry for CCSD(T)-level explorations to assess the relative stability of iso-energetic structures is non-trivial, as approximate methods exclusively prefer symmetric structures.  
Consequently, the true nature of the PES of 1AP remains elusive.

Given the complexity of the problem, an out-of-the-box approach is necessary to investigate the structural preferences of APs. 
We examine symmetry-lowering in APs from the highest point group symmetry, permitted by their composition and connectivities, to their maximum-order subgroups. 
To probe for symmetry lowering and the variation of STG on a lower-dimensional potential energy surface, we conducted a
diagnostic procedure, which we denote as Jahn--Teller--Hund's (JTH) diagnostics. 
We determine the PES of APs using the CCSD(T) method,
which is more accurate than MP2, to probe for changes in qualitative aspects of the PES due to electron correlation effects. 
Further, we inspect the variation of the STG across the PES.

\section{Computational details \label{sec_comp}}
For all APs shown in FIG.~\ref{fig_systems}, we performed full geometry optimizations 
by constraining the point group symmetries and relaxing the internal coordinates defined through the Z-matrix representation; see SI for the definitions of Z-matrices. The main goal of our study is to verify whether the high-symmetric structures of the APs shown in FIG.~\ref{fig_systems} correspond to minima on the PES. Hence, we have separately performed full geometry optimizations at the CCSD(T)/cc-pVTZ level for high-symmetric and low-symmetric structures. For all six APs, the high-symmetric structures have been reported in an earlier study\cite{loos2023heptazine}. In cases where the high-symmetric structures are saddle points on the PES, they cannot be used as a starting point to determine the low-symmetric structures because, at the saddle points, the forces on the atoms are zero. Hence, we select initial geometries of low-symmetric structures identified in JTH diagnostic analysis performed with the smaller basis set cc-pVDZ.

\begin{figure*}[!htpb]
\centering
\includegraphics[width=\linewidth]{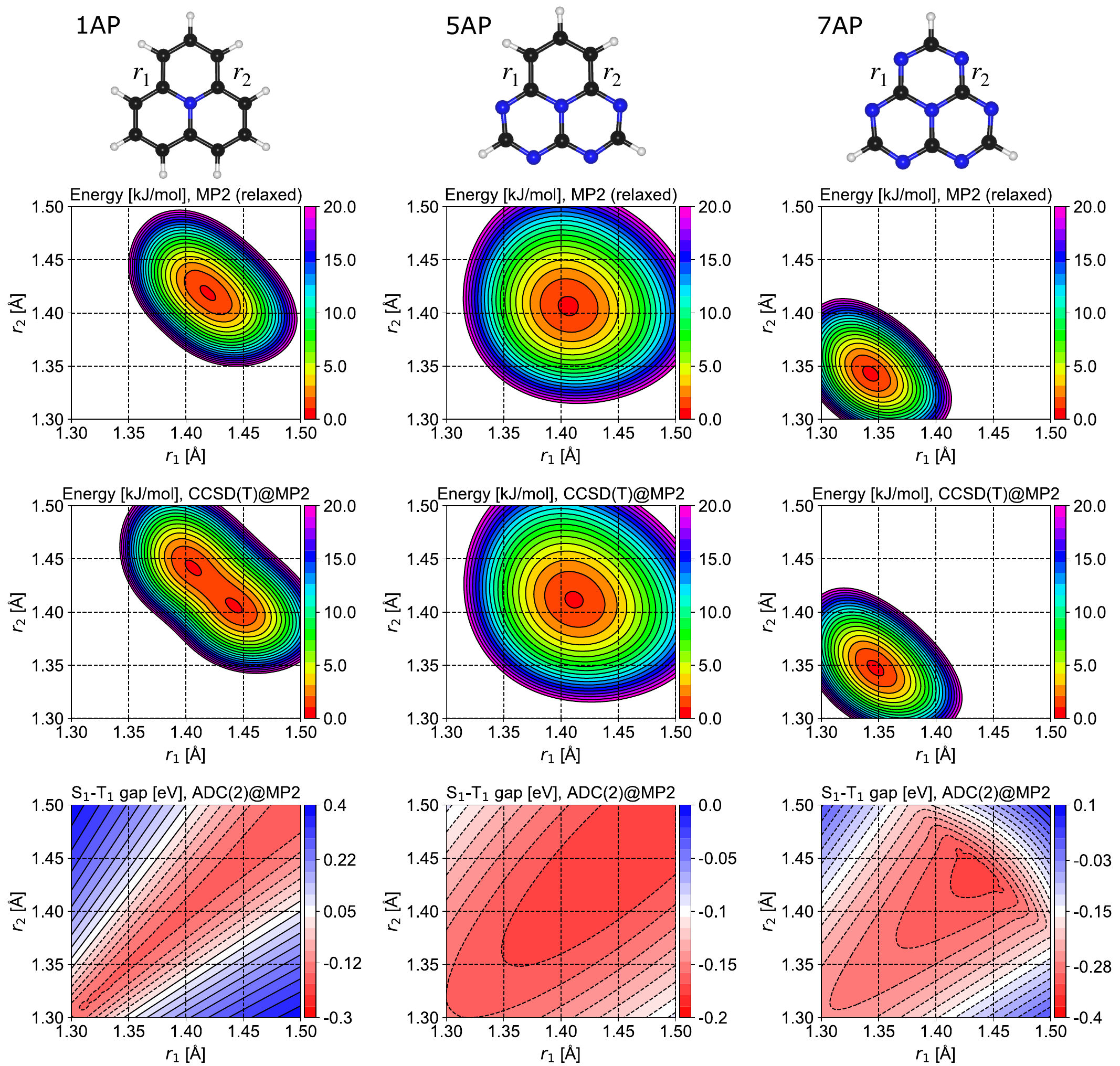}
\caption{
Contour plots of two-dimensional potential energy surfaces of the ground state and corresponding S$_1$-T$_1$ gaps for 1AP, 5AP, and 7AP. 
MP2 (relaxed) indicates that each point on the surface was calculated through constrained geometry optimization 
at the MP2-level for fixed values of
$r_1\in[1.30\,{\mathrm \AA}, 1.50\,{\mathrm\AA}]$ and 
$r_2\in[1.30\,{\mathrm \AA}, r_1]$, 
in steps of 0.01 \AA. 
Discrete points were interpolated as a surface using cubic functions.
The point group symmetries were constrained to \cthreeh\ for 1AP and 7AP, and \cs\ 
for 5AP; along $r_1=r_2$, the symmetries are \dthreeh and \ctwov, respectively.
CCSD(T)@MP2 indicates the ground state energy, while ADC(2)@MP2 indicates S$_1$-T$_1$ gap 
calculated on geometries optimized with MP2 at each point. 
 The ground state energy (in kJ/mol) and the S$_1$-T$_1$ gap (in eV) are color-coded by rainbow and red-to-blue color bars, respectively. Energies above 20 kJ/mol are not shown.
In all calculations, cc-pVDZ basis set was used.
\label{fig_PES_STG_1}}
\end{figure*}

\begin{figure*}[!htpb]
\centering
\includegraphics[width=\linewidth]{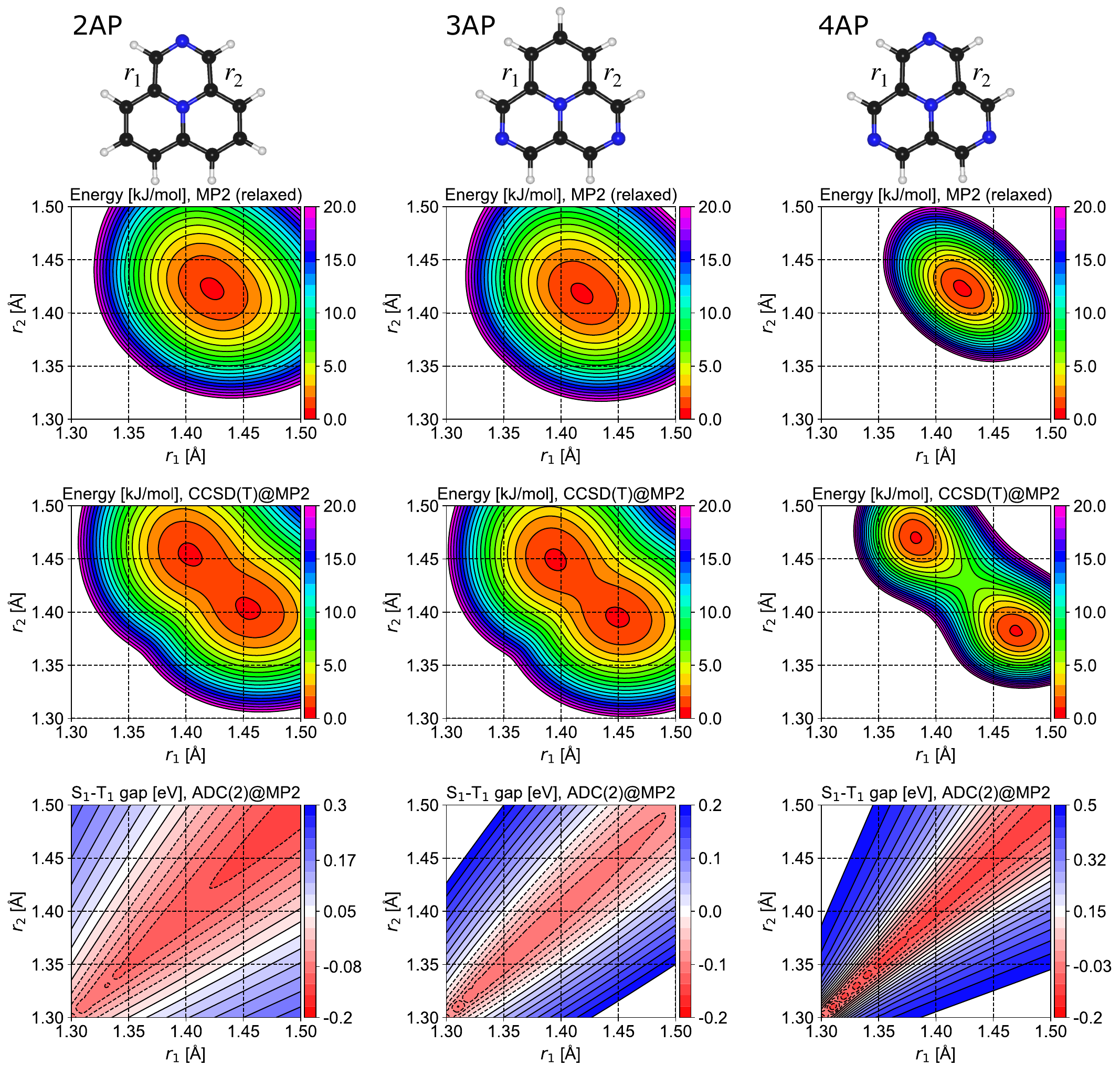}
\caption{
Contour plots of two-dimensional potential energy surfaces and corresponding S$_1$-T$_1$ gaps for 2AP, 3AP, and 4AP. 
The point group symmetries were constrained to 
\cthreeh\ for 4AP, 
\cs\ for 2AP and 
3AP; along $r_1=r_2$, the symmetries are \dthreeh and \ctwov, respectively. See the caption
of FIG.~\ref{fig_PES_STG_1} for further details.
\label{fig_PES_2AP2_3AP2_4AP2}}
\end{figure*}

As the first step of JTH diagnostics, MP2/cc-pVDZ-level constrained geometry optimizations were performed using Z-matrices by fixing 
two interatomic distances ($r_1$ and $r_2$ in FIGs.~\ref{fig_PES_STG_1} and \ref{fig_PES_2AP2_3AP2_4AP2}) at $r_1\in[1.30\,{\mathrm \AA}, 1.50\,{\mathrm\AA}]$  and $r_2\in[1.30\,{\mathrm \AA}, r_1]$, in steps of 0.01 \AA. For plotting the resulting two-dimensional PES, discrete points were interpolated using cubic functions as implemented in the {\tt matplotlib} module in Python. To probe the effect of electron correlation and the variation of STG across the PES, single-point calculations of the ground and excited state energies were calculated using the CCSD(T) and ADC(2) methods, respectively, in combination with the cc-pVDZ basis set. 

Structures corresponding to the lowest CCSD(T)/cc-pVDZ energy in the JTH diagnostic plots were selected as initial structures for full optimization. For 1AP, 2AP, 3AP, and 4AP, these structures correspond to the low-symmetry point groups \cthreeh, \cs, \cs, and \cthreeh, respectively, indicating the likelihood of symmetry lowering. For the same systems, minimum energy structures along the $r_1=r_2$ line were selected as initial geometries for full optimization of high-symmetry point groups. For 5AP and 7AP, CCSD(T)/cc-pVDZ energies did not indicate symmetry lowering. Hence, geometry optimizations were performed only for the high-symmetry structures (\ctwov and \dthreeh for 5AP and 7AP, respectively).

For all molecules, equilibrium geometries were also determined using the DFT methods, 
B3LYP\cite{stephens1994ab} and $\omega$B97X-D\cite{chai2008long}, 
along with the wavefunction methods MP2, CCSD and CCSD(T) in combination with the
cc-pVTZ basis set. 
Harmonic frequency analyses were performed with DFT and MP2 methods to 
confirm if the structures were minima or saddle points on their 
respective potential energy surfaces.
DFT-level geometry optimizations were conducted using Gaussian (version~16~C.01)\cite{frisch2016gaussianshort}.
MP2, CCSD, and CCSD(T) geometry optimizations were performed using Molpro (version~2015.1)\cite{werner2015molpro}.
For determining STGs, S$_1$ and T$_1$ energies were calculated using the excited states method, ADC(2), employing the resolution-of-identity (RI) approximation\cite{vahtras1993integral, kendall1997impact}
as implemented in QChem (version~6.0.2)\cite{krylov2013q} at the cc-pVTZ basis set.
All calculations based on the correlated wavefunction methods MP2, CCSD, CCSD(T), and ADC(2) were performed with the frozen-core approximation.
Complete basis set (CBS) extrapolation was performed 
using the two-point formula that uses CCSD(T) energies determined 
with the cc-pVTZ and cc-pVQZ basis sets. Hartree--Fock (HF) 
energy from the cc-pVQZ basis set is taken as the reference 
for the CBS estimate, while the correlation energy is extrapolated using the formula
$E_{\rm corr}^n=E_{\rm corr}^{\rm CBS}+\alpha n^{-3}$, where $n$ is the cardinal number of the basis set. 
Normal mode and nucleus-independent chemical shifts (NICS)\cite{schleyer1996nucleus} analyses were conducted with the $\omega$B97X-D3 method and the cc-pVTZ basis set using Orca (version~5.0.4)\cite{neese2012orca, neese2018software}. 
For both analyses, minimum energy geometries were calculated using 
$\omega$B97X-D/cc-pVTZ were utilized.
The NICS value was calculated as the negative of the magnetic shielding of a ghost atom
placed 1 \AA{} above the centroid of each ring
to determine isotropic and out-of-plane NICS, 
denoted NICS(1)$_{\rm iso}$ and NICS(1)$_{\rm zz}$, respectively.  
In all calculations performed with Orca, we used the resolution-of-identity (RI) approximation\cite{vahtras1993integral,kendall1997impact} and 
the `chain-of-spheres' (COS) algorithm for exchange integrals (RIJCOSX).

\section{Results and Discussion\label{sec_results}}
\subsection{Jahn--Teller--Hund diagnostic plots}
We begin our analysis with the two-dimensional PESs of 1AP, 5AP, and 7AP, obtained through constrained geometry optimizations by fixing
two interatomic distances related by two-fold rotation, see
FIG.~\ref{fig_PES_STG_1}. 
Using the Z-matrix representation, we constrained the molecules to be in the \cthreeh (for 1AP and 7AP) and \cs (for 5AP) subspaces. 
MP2-level ground state electronic energy surface shows prominent single-well behavior for all three systems with
1AP, 5AP, and 7AP preferring \dthreeh,  \ctwov, and \dthreeh structures 
with $r_1=r_2=1.42$ \AA{}, 1.41 \AA, and 1.40 \AA, respectively. 
Subsequent MP2/cc-pVTZ-level geometry optimizations and 
harmonic vibrational frequency analysis 
described these symmetric structures as minima on the PESs.

CCSD(T) energies calculated in a single-point fashion identify two equivalent minima of 1AP 
corresponding to \cthreeh symmetry at
$\lbrace r_1=1.44~{\rm \AA},\, r_2=1.41~{\rm \AA} \rbrace$ and $\lbrace r_1=1.41~{\rm \AA},\, r_2=1.44~{\rm \AA} \rbrace$, 
see  FIG.~\ref{fig_PES_STG_1}.
The \dthreeh structure predicted as a minimum by MP2 shows characteristics of a saddle point at the CCSD(T)-level. 
It is important to note that the transition state structure is more sensitive to electron correlation effects and 
MP2 has a tendency to overestimate the importance of double excitations for transition state structures\cite{scuseria1990concerted}.
Another classic example is cyclobutyne, 
which MP2 predicts as a minimum but is found to be a first-order saddle point
by the more accurate CCSD(T) and CCSDT methods\cite{sun2019cyclobutyne}.
As the HOMO-LUMO gap increases upon substitution of N in the electron-rich sites of 1AP---with values of about $6/8/10$ eV for 1AP/5AP/7AP, respectively---electron correlation effects become less relevant. 
Hence, the PESs of 5AP and 7AP remain similar at both MP2 and CCSD(T) levels.
PES plots calculated with the DFT methods, B3LYP and 
$\omega$B97X-D, are on display in Figure~S1 in the 
Supplementary Information (SI). 
Both methods forecast single-well type PES for 1AP, 5AP, and 7AP.

In general, for probing the inverted nature of STGs of APs, correlated excited state methods such as ADC(2) or CC2 have provided good estimates\cite{tuvckova2022origin,loos2023heptazine}.  
For ten APs, the mean unsigned errors in the STGs predicted by ADC(2) and CC2 were reported to be 0.033 and 0.037 eV\cite{loos2023heptazine}. However, EOM-CCSD with similar computational cost as ADC(2) and CC2 has been shown to have a slightly larger error of 0.081 eV\cite{loos2023heptazine}. For 7AP, the similarity-transformed EOM-CCSD method has predicted a too-negative STG of $-$0.66 eV\cite{ghosh2022origin}, compared to its TBE value of $-$0.219 eV\cite{loos2023heptazine}.
STGs calculated on the PES with the ADC(2) method and cc-pVDZ basis set identifies regions where  
S$_1$ is lower in energy than T$_1$, see FIG.~\ref{fig_PES_STG_1}. 
For all three molecules, symmetric structures exhibit inverted STG. Individual plots of S$_1$ and T$_1$ relative to S$_0$ are on display in Figure~S2, where one notices that the S$_1$ energy drops more rapidly when deviating from the diagonal line (corresponding to symmetric structures) compared to the T$_1$ energy. 
Overall, the JTH diagnostics on a low-dimensional PES hints at a symmetry-lowered structure for 1AP with an STG of diminished magnitude while suggesting symmetric structures for 5AP and 7AP with negative STG in agreement with experimental measurements\cite{ehrmaier2019singlet,wilson2024spectroscopic}. 
For in-plane distortions from the equilibrium geometry along the two interatomic distances, 5AP shows robust Hund's rule violation compared to 7AP. For the latter, FIG.~\ref{fig_PES_STG_1} shows regions with positive STG. 

FIG~\ref{fig_PES_2AP2_3AP2_4AP2} presents
the JTH diagnostic plots for 2AP, 3AP, and 4AP. 
MP2 describes these systems as prominent single-well structures of maximum symmetry (\ctwov for 2AP and 3AP; \dthreeh for 4AP). 
MP2/cc-pVTZ geometry refinements and frequency analysis indicated all three symmetric structures to be minima. 
CCSD(T) energies show a trend contrasting with MP2-level description for all three molecules; in all cases, the high-symmetry structure turns out to be the saddle point connecting low-symmetry structures belonging to the largest subgroup (\cs for 2AP and 3AP; \cthreeh for 4AP ). As seen in FIG~\ref{fig_PES_2AP2_3AP2_4AP2}, the barrier increases 
with the number of N atoms on charge-destabilizing positions (\ie LUMO sites of 1AP, see FIG.\ref{fig_systems}). 
STG plots of all three molecules indicate a profile similar to the one noted for 1AP in FIG.~\ref{fig_PES_STG_1}. The STG is negative along the $r_1=r_2$ line containing 
the high-symmetry structures, while for the distorted structures, the plot suggests 
positive STGs. 
CCSD(T)/cc-pVTZ geometries were determined using the coordinates of the distorted structures inferred from FIG~\ref{fig_PES_2AP2_3AP2_4AP2} as 
the starting point. Subsequent ADC(2)/aug-cc-pVTZ excited state calculations for 2AP, 3AP, and 4AP
resulted in STGs of +0.125 eV, $+$0.274 eV, and +0.392 eV, respectively (see Table~\ref{tab_table1}).


Figure~S1 shows the PESs of 2AP, 3AP, and 4AP generated using the DFT methods B3LYP and $\omega$B97X-D. While B3LYP predicts 2AP and 3AP to be single-wells,  for 4AP, the method suggests a slight distortion with $r_1$ and $r_2$ differing by 0.03 \AA{}. Frequency analysis confirmed the \dthreeh and \ctwov structures of 2AP and 3AP to be
minima at the B3LYP/cc-pVTZ-level, while the symmetric structure of 4AP showed one imaginary frequency (494$i$ cm$^{-1}$, $A_2^\prime$). 
The long-range corrected DFT method, $\omega$B97X-D, has been shown to outperform B3LYP in determining minimum energy structures of unusual covalent bond connectivities\cite{senthil2021troubleshooting}. For 2AP, $\omega$B97X-D shows a weak distortion in agreement with CCSD(T). At the same time, the method indicates the symmetric geometries of 2AP, 3AP, and 4AP to be saddle points, suggesting the possibility of locating the low-symmetry minima with less expensive DFT modeling (see Figure~S1).

\subsection{Geometries and Singlet-Triplet gaps}
\begin{figure}[!htpb]
\centering
\includegraphics[width=\linewidth]{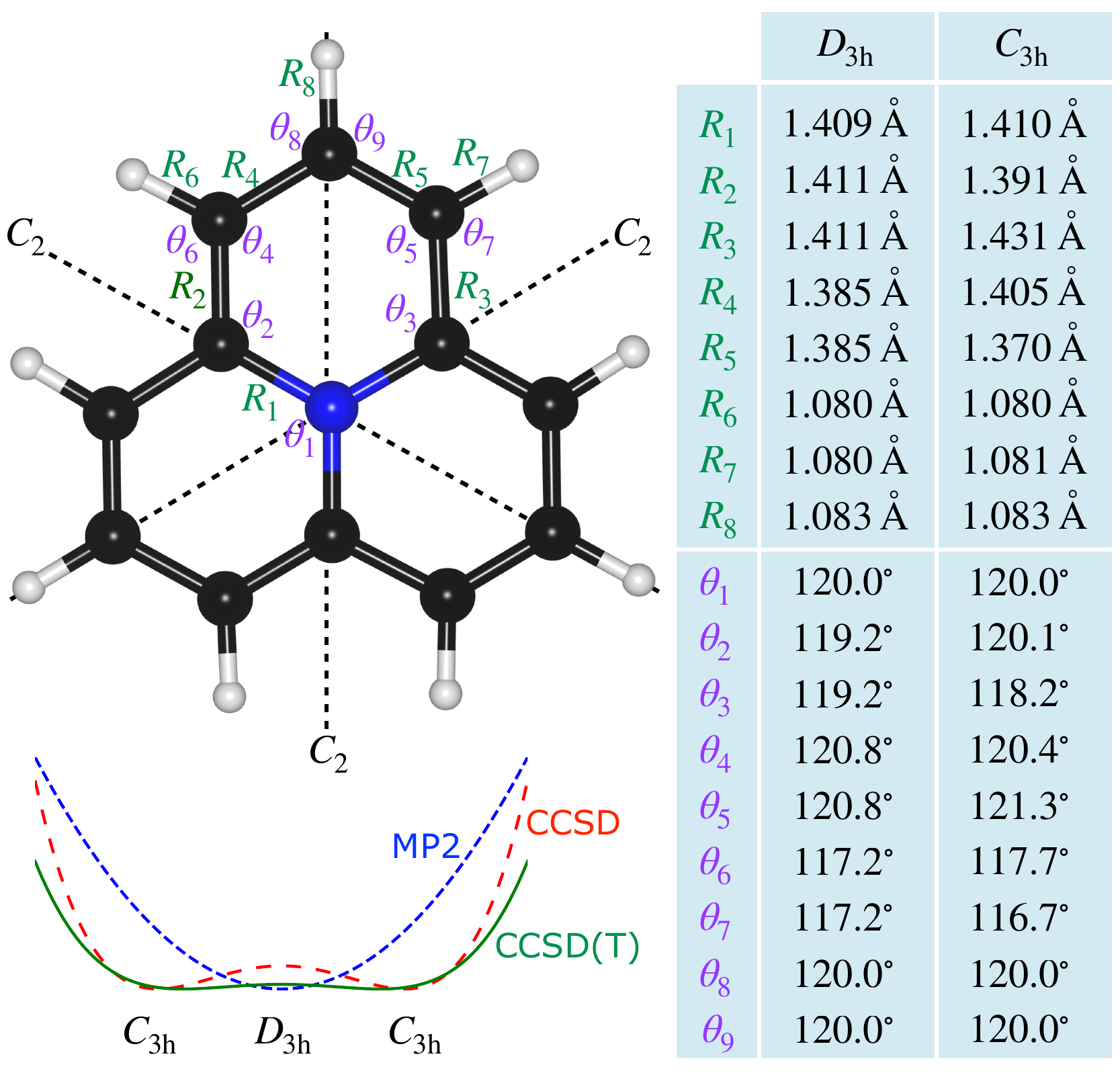}
\caption{
Equilibrium geometry parameters of 1AP 
in \dthreeh (saddle point) and \cthreeh (minimum) configurations calculated with the CCSD(T)/cc-pVTZ method. 
Dashed lines show two-fold rotation axes, perpendicular to the principal (three-fold rotation) axis, present in \dthreeh. 
The schematic PES shows that MP2 describes 1AP as a single-well with the \dthreeh  
structure to be the minimum, 
while CCSD and CCSD(T) suggest a double-well behavior with the lower symmetric \cthreeh structure as the minimum. 
At the CCSD(T)-level, the barrier to automerization is vanishingly small, indicating the 
vibrational ground state's wavefunction is strongly delocalized over the equivalent \cthreeh minima.
}
\label{fig_1AP_geometries}
\end{figure}

Table~\ref{tab_table1} presents STGs of the selected APs shown in FIG.~\ref{fig_systems} 
calculated with ADC(2)/aug-cc-pVTZ using equilibrium geometries determined using the accurate CCSD(T)/cc-pVTZ method.
Results based on geometries from other methods are available in the SI (see Tables~S1, S2, and S3 for 1AP, 5AP, and 7AP, respectively; Table~S4 for 2AP, 3AP, and 4AP). 
FIG.~\ref{fig_1AP_geometries} shows the corresponding equilibrium geometry parameters for both the structures at the CCSD(T)/cc-pVTZ-level. 
For the \dthreeh structure of 1AP, our ADC(2) value of the STG=$-$0.142 eV, deviates slightly 
from the previously obtained\cite{loos2023heptazine} TBE of $-$0.131 eV.
Using the basis set limit values of CCSD(T) energies,
we find the \cthreeh structure of 1AP to be lower in energy than the \dthreeh structure by only 0.3 kJ/mol.

Had the symmetry lowering been predicted by MP2 or DFT methods but not at the more accurate CCSD(T)-level, then the distortion should be considered artifactual\cite{davidson1983symmetry}. 
According to the accurate CCSD(T)-level description, the actual PES of 1AP should be thought of as a very shallow double well permitting
strong carbon tunneling\cite{zuev2003carbon}, which according to quantum mechanics has to be interpreted as 
two equivalent \cthreeh structures undergoing
rapid automerization. Furthermore, as pointed out earlier by 
Leupin \etal the nuclear magnetic resonance (NMR) 
spectrum of 1AP does not rule out the possibility of a rapid
isodynamic equilibrium between two \cthreeh minima\cite{leupin1980low}, which aligns with its very low barrier
reported in this study.

\begin{table}[!htpb]
\begin{threeparttable}
\centering
\caption{
For the six azaphenalenes shown in FIG.~\ref{fig_systems}, vertical excitation
energies of the S$_1$ and T$_1$ states with respect to the ground state, S$_0$,
along with the singlet-triplet gap (S$_1$-T$_1$) are given in eV.  
The excited state energies were calculated using the ADC(2)/aug-cc-pVTZ
method using geometries determined with the CCSD(T)/cc-pVTZ method. The barrier for automerization, $E^\ddagger$ (in kJ/mol), determined 
using two-point CBS extrapolation of CCSD(T) energies with 
cc-pVTZ and cc-pVQZ basis sets, is stated for the symmetric saddle point.
Results from other studies are included for 
comparison; 0-0 indicates adiabatic transition energies 
accounting for the energy of the vibrational ground state.
}
\label{tab_table1}
\addtolength{\tabcolsep}{1.2pt}
\begin{tabular}[t]{lll lll }
\hline
\multicolumn{1}{l}{System}   &
\multicolumn{1}{l}{$E^\ddagger$ } & 
\multicolumn{1}{l}{S$_1$} &
\multicolumn{1}{l}{T$_1$} &
\multicolumn{1}{l}{S$_1$-T$_1$} & 
\multicolumn{1}{l}{Source} \\
\hline 
1AP (\dthreeh)&  0.3    & 0.979  & 1.121 & $-$0.142   & this work   \\ 
1AP (\cthreeh)&   & 1.130  & 1.190 & $-$0.060      & this work  \\ 
1AP           &      & 1.047 & 1.180 & $-$0.133     & ADC(2)\cite{tuvckova2022origin}  \\ 
1AP           &      & 0.992 & 1.068 & $-$0.076     & 0-0, ADC(2)\cite{tuvckova2022origin}  \\ 
1AP (\dthreeh)&      & 0.979  & 1.110 & $-$0.131   & TBE\cite{loos2023heptazine}\\
1AP           &      & 0.97   & 0.93  & +0.04      & exp.\cite{leupin1980low}$^a$\\
\multicolumn{6}{l}{}\\
5AP (\ctwov)  &      &  2.128      &  2.274     & $-$0.147  & this work   \\ 
5AP           &      & 2.231 & 2.365 & $-$0.134     & ADC(2)\cite{tuvckova2022origin}  \\ 
5AP           &      & 1.971 & 2.056 & $-$0.085     & 0-0, ADC(2)\cite{tuvckova2022origin}  \\ 
5AP (\ctwov)  &      &  2.177 & 2.296 & $-$0.119   & TBE\cite{loos2023heptazine}\\
5AP           &      & 1.957  & 2.003 & $-$0.047   & 0-0, exp.\cite{wilson2024spectroscopic}\\
\multicolumn{6}{l}{}\\
7AP (\dthreeh)&      &  2.636      &  2.889     & $-$0.253  & this work   \\ 
7AP           &      & 2.756 & 2.998 & $-$0.242     & ADC(2)\cite{tuvckova2022origin}  \\ 
7AP           &      & 2.512 & 2.618 & $-$0.106     & 0-0, ADC(2)\cite{tuvckova2022origin}  \\ 
7AP (\dthreeh)&      & 2.717 & 2.936 & $-$0.219    & TBE\cite{loos2023heptazine}\\
7AP           &      &        &       & $\quad<$0       & exp.\cite{ehrmaier2019singlet}\\
\multicolumn{6}{l}{}\\
2AP (\ctwov)  &   2.1   &    0.838    &  0.941     & $-$0.102  & this work   \\ 
2AP (\cs)     &   &   1.246     &  1.121     & $+$0.125  & this work  \\ 
2AP (\ctwov)  &      &  0.833      & 0.904      & $-$0.071   & TBE\cite{loos2023heptazine}\\
\multicolumn{6}{l}{}\\
3AP (\ctwov)  &   5.3   &    0.689    &  0.768     & $-$0.079  & this work   \\ 
3AP (\cs)     &   &   1.335     &   1.061    & $+$0.274  & this work  \\ 
3AP (\ctwov)  &      & 0.693       & 0.735      & $-$0.042   & TBE\cite{loos2023heptazine}\\
\multicolumn{6}{l}{}\\
4AP (\dthreeh)&    11.1  &   0.550     &  0.623     & $-$0.073  & this work   \\ 
4AP (\cthreeh)&   &  1.409     &  1.017     & $+$0.392  & this work  \\ 
4AP (\dthreeh)&      &  0.554      &    0.583   & $-$0.029   & TBE\cite{loos2023heptazine}\\
\hline
\end{tabular}
 \begin{tablenotes}
  \item \footnotesize{$^a$ Using S$_1$=0.78 $\mu$m$^{-1}$ and T$_1$=0.75 $\mu$m$^{-1}$ 
  from \RRef{leupin1980low} multiplied by 1.2398 eV/$\mu$m$^{-1}$.}
  \end{tablenotes}
\end{threeparttable}
\end{table}%

Rigorous confirmation of the true sign of the STG of 1AP requires consideration of 
high-level effects such as beyond-ADC(2) treatment of excited state and adiabatic effects along with the symmetry lowering effect. The ADC(2) level STG of the \dthreeh structure of 1AP is $-$0.142 eV determined using vertical excitation energies. Post-ADC(2), high-level corrections can be estimated as $+0.009$ eV by comparing the STG of the \dthreeh structure from ADC(2) ($-$0.142 eV) with the TBE ($-$0.131 eV) from \RRef{loos2023heptazine}, see Table~\ref{tab_table1}. Furthermore, for the \dthreeh structure of 1AP, the vertical STG underestimates 
the adiabatic transition energy (0-0), which accounts for geometric effects in S$_1$ and T$_1$ along with 
the zero-point vibrational energy, by +0.057 eV at the ADC(2)-level\cite{tuvckova2022origin}. Using these two corrections, one can estimate the true STG of the \dthreeh structure to be $-$0.076 eV. Further, for  
1AP in the \cthreeh structure, the ADC(2)/aug-cc-pVTZ value of STG is $-0.06$ eV shifting the \dthreeh value by +0.082 eV. Hence, one can estimate the 0-0 STG of the \cthreeh structure to be $+0.006$ eV 
explaining the small positive experimental STG of +0.04 eV\cite{leupin1980low}. 
Given the vanishing barrier of 0.3 kJ/mol 
for automerization, one can expect the wavefunction of the 
vibrational ground state to span double well broadly. Hence, when the 0-0 STG of 1AP is weighted by 
the square of this wavefunction, one can expect STG to be vanishingly small, as in the \cthreeh structure.

We performed separate full geometry optimizations for high-symmetry
and low-symmetry geometries of 5AP and 7AP, and found no evidence for symmetry lowering via
in-plane distortions. 
For 5AP and 7AP,  STGs calculated with ADC(2)/aug-cc-pVTZ
based on geometries determined with CCSD(T)/cc-pVTZ  
show deviations of about 0.03 eV from TBEs. 
The structural stability of 5AP and 7AP, confirmed by the CCSD(T)-level analysis, indicates that both molecules violate Hund's rule in their minimum energy geometry with high-symmetry. 
From previously reported\cite{tuvckova2022origin} ADC(2)-level vertical 
and 0-0 values of STG, one can correct the TBE of the vertical STG of 5AP from \RRef{loos2023heptazine} to be $-$0.07 eV, agreeing with the experimental value of
$-0.047$ eV\cite{wilson2024spectroscopic}. As the precise value
of the STG of 7AP is unknown, the TBE values corrected for
0-0 excitation amounts to $-0.083$ eV, which is lower than $-$0.011 eV from the fluorescence measurements\cite{aizawa2022delayed} 
of a substituted 7AP.


We found the NICS(1)$_{\rm iso}$ values of 5AP and 7AP to have very small magnitudes, indicating these two systems to be non-aromatic with localized charge density. The other four systems in their symmetric structure exhibit large positive NICS(1)$_{\rm iso}$ values 1 \AA{} above the centroid of their rings indicating anti-aromaticity\cite{karadakov2008ground}, for the symmetric geometries determined at the $\omega$B97X-D/cc-pVTZ-level, in the order: 1AP $<$ 2AP $<$ 3AP $<$ 4AP (see Figure S4). 

To understand whether anti-aromaticity increases the multi-reference character of their ground state, we inspect the $T_1$-diagnostic metric, which is the Frobenius norm of the single substitution amplitudes vector ($t_1$) obtained from the closed-shell CCSD wave function divided by the square root of the number of correlated electrons\cite{wang2015multireference}. A magnitude of $T_1$ less than 0.02 indicates that the ground state is predominantly of a single-reference nature for which the single-reference method, CCSD(T), which we have applied for geometry optimization, is appropriate. For fully optimized structures at the CCSD(T)/cc-pVTZ level, we found the  $T_1$ values to be:  
0.017 (1AP: \dthreeh \& \cthreeh), 
0.017 (2AP: \ctwov), 
0.015 (2AP: \cs), 
0.016 (3AP: \ctwov), 
0.015 (3AP: \cs), 
0.016 (4AP: \dthreeh \& \cthreeh), indicating single-reference nature. 
Interestingly, the $T_1$ values for 5AP (\ctwov) and 7AP (\dthreeh) were close to the threshold, 0.019 and 0.020, respectively, indicating the need to investigate these systems with multi-reference methods. Furthermore, as the systems with the small NICS(1)$_{\rm iso}$ values---5AP and 7AP---show larger $T_1$ values, for azaphenalenes, one can conclude that anti-aromaticity does not imply a multi-reference character of the ground state.


\subsection{Jahn--Teller Analysis}
We rationalize the symmetry lowering in APs
as a vibronically driven process on the basis of the Jahn--Teller (JT) theorem\cite{jahn1937stability,bader1962vibrationally,pearson1975concerning} . 
Specifically, we verify if the symmetry 
of the vibrational normal mode corresponding to the distortion, as seen in DFT results, agrees with the JT theorem. 
The electronic ground state energy, $E_0(q)$, is expanded as a truncated Taylor series in the
distortion coordinate, $q$:
\begin{eqnarray}
    E_0(q) & = & E_0 + \langle 0 | \hat{H}^\prime  | 0 \rangle q + 
    \frac{1}{2} k q^2,
     \label{eq_JT_1}
\end{eqnarray}
where $E_0$ corresponds to the ground state energy in the undistorted equilibrium geometry (at $q=0$), and
$\hat{H}^\prime$ is the first derivative of the electronic Hamiltonian with respect to $q$. 
At $q=0$, the coefficient in the first-order JT contribution, $\langle 0 | \hat{H}^\prime  | 0 \rangle$,
vanishes for APs; however, this term 
can be non-zero for systems with degenerate electronic states.
For pseudo-degenerate (\ie near degenerate)
electronic states, the second-order term in $q^2$ can be non-zero, where 
$k$ is the total force constant that includes contributions from vibronic coupling\cite{nakajima1982geometrical}
\begin{eqnarray}
    k & = &   \langle 0 |\hat{H}^{\prime\prime}| 0 \rangle - 
  2  \sum_{n \ne 0} \frac{ |\langle 0 |\hat{H}^{\prime}| n \rangle|^2 }{E_n - E_0}.
   \label{eq_JT_2}
\end{eqnarray}
Here $\hat{H}^{\prime\prime}$ is second derivative of the 
electronic Hamiltonian with respect to $q$, and
$E_n$ is the energy of the $n$-th electronic state at $q=0$.

The second-order term gives rise to the pseudo-JT contribution capable of stabilizing the system upon distortion from $q=0$, 
provided there is a low-lying state, $n$, of correct symmetry such that $\langle 0 |\hat{H}^{\prime}| n \rangle \ne 0$. For this, the product of the corresponding irreducible representations, $\Gamma_0 \times \Gamma_q \times \Gamma_n$, should contain the totally symmetric representation of the point group corresponding to the symmetric structure. If this group is Abelian, then the condition is $\Gamma_q=\Gamma_0 \times \Gamma_n$.
For 1AP and 4AP, the totally symmetric irreducible representation is $A_1^\prime$ of \dthreeh, and for 2AP and 3AP, it is $A_1$ of \ctwov. We write the irreducible representations in small case alphabets when denoting the MOs and use large cases for the states and vibrational modes. 

 Assuming that there is only one low-lying excited state, $| 1\rangle$, coupling with the ground state, 
 we arrive at
\begin{eqnarray}
    k & = &   \langle 0 |\hat{H}^{\prime\prime}| 0 \rangle - 
  2   \frac{ |\langle 0 |\hat{H}^{\prime}| 1 \rangle|^2 }{E_1 - E_0} 
       =  k_0 - k_{0,1}.
      \label{eq_JT_3}
\end{eqnarray}
Here, $k_0\ge0$ is the force constant of the symmetric structure in the absence of vibronic coupling, while the condition for distortion is $k<0$.  
The second term of Eq.~\ref{eq_JT_3}, $k_{0,1}$, cannot be negative, as it is a ratio of two positive numbers. 
Hence, symmetry-lowering is prominent when  $k_0<k_{0,1}$, which requires a significantly large
$|\langle 0 |\hat{H}^{\prime}| 1 \rangle|^2$, small $E_1-E_0$ or both. This explains why symmetry lowering is not observed in 5AP and 7AP with fairly large S$_0$$\rightarrow$S$_1$ excitation energies of about 2.1, and 2.6 eV 
(see TABLE~\ref{tab_table1}). On the other hand, the ADC(2)/cc-pVTZ excitation energy of S$_1$ is 1.0 eV in 1AP (\dthreeh), which decreases to 0.8, 0.7, and 0.6 eV in 2AP (\ctwov), 3AP (\ctwov), and 4AP (\dthreeh), respectively. Overall, our analysis suggests that inverted-STG candidates with small values of ${\rm S}_0 \rightarrow {\rm S}_1$ may exhibit strong structural distortions warranting more detailed analysis such as the JTH diagnostics presented in this study.

The symmetry of the ground electronic state of the APs,
in their high-symmetry geometry,
correspond to the totally symmetric irreducible representations: 
$A_1$ of \ctwov  and $A_1^\prime$ of \dthreeh. 
If the irreducible representation of $|1\rangle$ is non-degenerate, the product of the irreducible representations of $\hat{H}^{\prime}$ and $| 1 \rangle$ should be totally symmetric, which is possible only when $q$ and $| 1 \rangle$ belong to the same irreducible representation.
The character of the HOMO and LUMO in all APs are similar, see Figure~S3.
For molecules in \dthreeh point group, the symmetries are $a_1^{\prime\prime}$ and $a_2^{\prime\prime}$, respectively, while for those in \ctwov, the symmetries are $a_2$ and $b_1$, respectively. The symmetry of the excited state, $|1\rangle$, when considering a Slater-determinant form, is the product, $\Gamma_{\rm HOMO}\times\Gamma_{\rm LUMO}$. Hence, the symmetry
of the excited state under concern is $A_2^\prime$ for 1AP and 4AP (in \dthreeh) and $B_2$ for 2AP and 3AP (in \ctwov).

In our $\omega$B97X-D/cc-pVTZ calculations, the wavenumbers
of the distortion modes of 2AP, 3AP, and 4AP  are 
340$i$ cm$^{-1}$ ($B_2$), 
773$i$ cm$^{-1}$ ($B_2$), and
1113$i$ cm$^{-1}$ ($A_2^\prime$), respectively; see Table~S4 in the SI. 
The corresponding $k$ (in Eq.~\ref{eq_JT_3}) are 
$-$0.6402,
$-$3.8163, and 
$-$8.6171 mdyne/\AA, respectively. 
This trend shows how the distortion becomes stronger
with increasing N in the electron-deficient sites of 1AP.
While we have presented evidence for enhanced distortion with the lowering of
$E_1-E_0$, it is also important to understand the role of the coupling term $|\langle 0 | \hat{H^\prime}| 1 \rangle|^2$. One can 
heuristically estimate this quantity through two conditions: 
1) the expectation value of the second-derivative term is similar in both distorted and
undistorted geometries, 
2) in the low-symmetry (\ie distorted) structure, there is no vibronic coupling. 
These two conditions amount to 
\begin{eqnarray}
    k_{\rm high-sym.} = k_{\rm low-sym.} -  2   \frac{ |\langle 0 |\hat{H}^{\prime}| 1 \rangle|^2 }{E_1 - E_0}. 
    \label{eq_k_lowsymm}
\end{eqnarray} 
The meaning of $|\langle 0 |\hat{H}^{\prime}| 1 \rangle|$ is the magnitude of force due to vibronic coupling in the high-symmetry structure; this quantity will increase in magnitude, going from 2AP to 3AP to 4AP in the same order as the strength of the distortion. 
The corresponding $k_{\rm low-sym.}$ (in Eq.~\ref{eq_k_lowsymm}),
determined using frequency calculations at the minimum energy geometries, 
are 0.0454, 0.0479, and 0.0522 mdyne/\AA, respectively.

Hence, one can determine $|\langle 0 |\hat{H}^{\prime}| 1 \rangle|$ as
\begin{eqnarray}
    |\langle 0 |\hat{H}^{\prime}| 1 \rangle| = \sqrt{ \frac{ (k_{\rm low-sym.} - k_{\rm high-sym.}) (E_1 - E_0) }{2} }.
        \label{eq_k_lowsymm2}
\end{eqnarray}
resulting in 1.308, 2.906, and 4.029 eV/\AA{} for 2AP, 3AP, and 4AP, respectively. 
At the $\omega$B97XD level 1AP does not undergo distortion, hence in Eq.~\ref{eq_k_lowsymm}, $k_{\rm high-sym.} = k_{\rm low-sym.}$ resulting in
$|\langle 0 |\hat{H}^{\prime}| 1 \rangle|=0$. The increase in the strength of 
distortion in the order 2AP$<$3AP$<$4AP is a combined effect of decreasing S$_1-$S$_0$ gap, and increasing
$|\langle 0 |\hat{H}^{\prime}| 1 \rangle|$ in Eq.~\ref{eq_JT_3}.

The Jahn--Teller active normal mode corresponds to in-plane stretching, 
with symmetries $A_2^\prime$ (for \dthreeh $\rightarrow$ \cthreeh transition) and $B_2$ (for \ctwov $\rightarrow$ \cs transition). 
Upon symmetry lowering, the irreducible representation is $A^\prime$, which is the totally symmetric
representation in \cthreeh and \cs.
This observation follows the epikernel principle: preferred distortions 
of JT unstable molecules are directed towards the largest subgroup
such that the JT active mode in the subgroup is totally symmetric
\cite{ceulemans1984symmetry}. 

\subsection{Ground and excited state energies of 4AP along Jahn--Teller active normal mode}
\begin{figure}[!htpb]
\centering
\includegraphics[width=\linewidth]{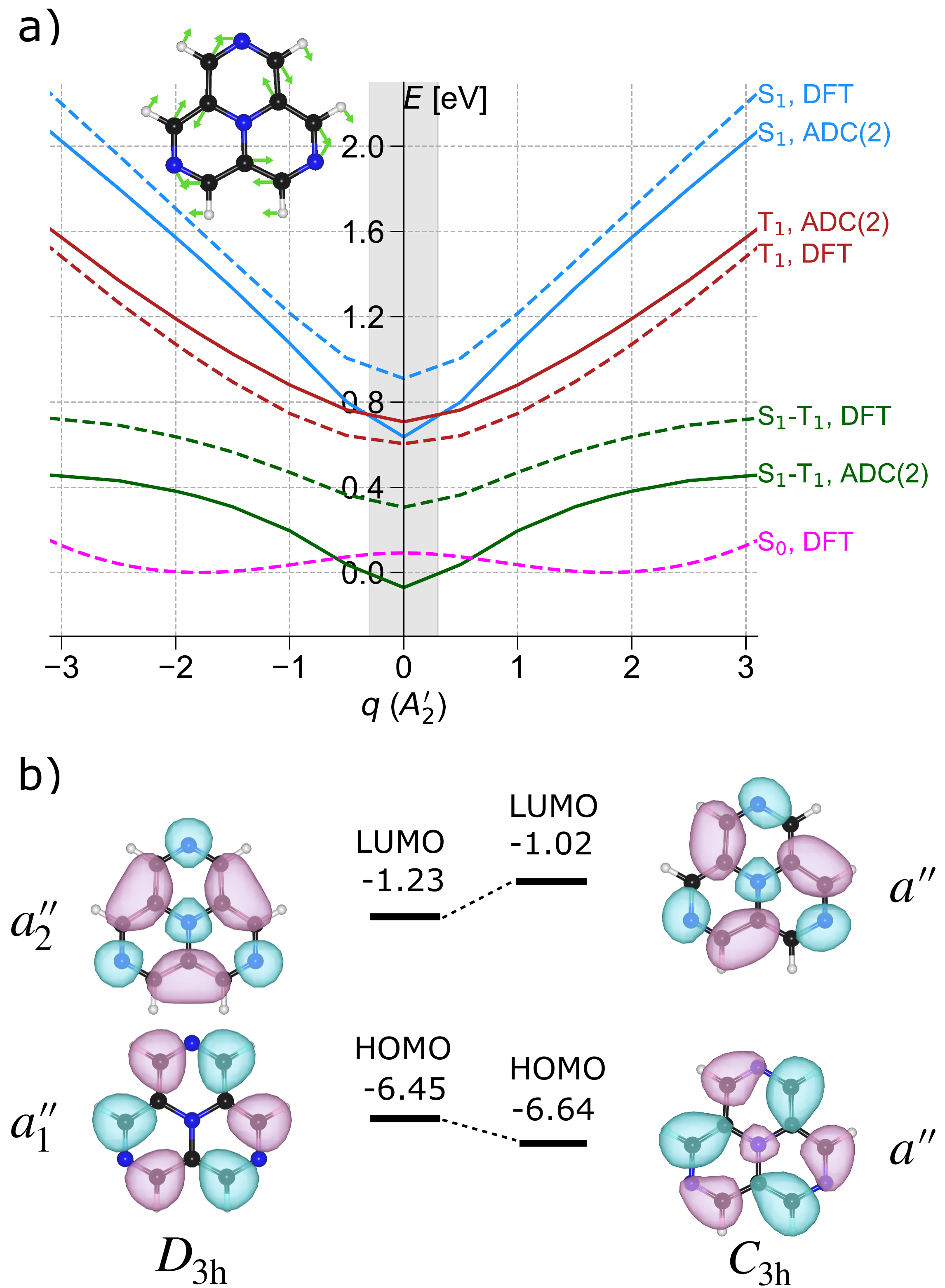}
\caption{
Pseudo-Jahn--Teller distortion in 4AP according to $\omega$B97X-D/cc-pVTZ modeling:
a) Potential energy profile of 4AP along the Jahn--Teller active dimensionless normal coordinate, $q$, 
of $A_2^\prime$ symmetry. 
The force constant matrix was computed at the \dthreeh geometry ($q=0$), a first-order saddle point with an imaginary frequency (1113$i$ cm$^{-1}$); the inset shows the normal mode. 
The DFT ground electronic state energy is 
denoted S$_0$ (pink, dashed) and the symmetry lowers to \cthreeh when $q\ne0$. 
The excited state energies, S$_1$ (blue), T$_1$ (red), and singlet-triplet gap, S$_1$-T$_1$, (green) determined using ADC(2)/cc-pVTZ and $\omega$B97X-D3/cc-pVTZ methods are represented by bold and dashed lines, respectively. The singlet-triplet gap is negative within the shaded region $-0.3 \le q \le 0.3$ at the ADC(2) level. 
b) Frontier MOs of 4AP in \dthreeh and \cthreeh structures showing the HOMO-LUMO gap increases due to better symmetry-allowed atomic orbital interactions in the \cthreeh structure.
}
\label{fig_normalmode}
\end{figure}
FIG.~\ref{fig_normalmode}a shows the ground and excited state energies of 4AP along the JT active mode of the \dthreeh configuration. The S$_0$ energy was determined at the DFT level, while the excited state energies were calculated using ADC(2) and DFT methods. 
The S$_0$ profile shows the double-well character at the DFT level, whereas, at the same DFT level, the S$_1$ and T$_1$ profiles are predicted as single wells. The excited state energies also show single-well character at the ADC(2) level. It is interesting to note that the ground state analogue of ADC(2), MP2, predicts the structure of 4AP to be \dthreeh in disagreement with CCSD(T). 
The notable difference between predictions of the excited state profiles by both methods is that according to ADC(2), S$_1$ has lower energy than T$_1$ at the \dthreeh configuration ($q=0$). The ADC(2)-level STG remains negative for slight distortions from the \dthreeh structure, while 
at $q\thickapprox$1.8 (corresponding to the \cthreeh minimum), the STG increases to +0.344 eV (see Table~S4). At the $\omega$B97X-D3 level, with the Tamm--Dancoff approximation, 
the STG is positive and somewhat uniform for all values of $q$. 
Overall, the DFT and ADC(2) profiles shown in FIG.~\ref{fig_normalmode}a suggest that a method capable of predicting the double-well profile for S$_0$ does not need to predict the negative sign of the STG and {\it vice versa}.

The main driving force for the pseudo-JT effect is the improved orbital interactions permitted by symmetry lowering. Analysis of the HOMO and LUMO plots of 4AP in \dthreeh and \cthreeh configurations reveals that the $a_1^{\prime\prime}$ and $a_2^{\prime\prime}$ MOs (HOMO and LUMO) transform as $a^{\prime\prime}$ upon symmetry lowering, see FIG.~\ref{fig_normalmode}b. A mild splitting of the HOMO-LUMO gap is noted upon distortion, imparting more stability to the ground state. Subsequently, symmetry lowering enhances the exchange interaction between HOMO and LUMO in the \cthreeh structure, 
resulting in a more stable T$_1$ state compared to S$_1$.

\section{Conclusions\label{sec_conc}}

In conclusion, we conducted a diagnostic analysis followed by full geometry optimizations to determine if the symmetric structures of APs are true minima on the PES and to investigate the dependence of their STG on structural distortions. 
Out of six APs studied here, the \dthreeh structure of the prototypical molecule, cyclazine (1AP), is predicted to be a saddle point by CCSD(T) with the correct minimum identified to have the \cthreeh symmetry. 
The mildly anti-aromatic nature of the ring-current in the symmetric form suggests the possibility of symmetry-lowering as in the case of cyclobutadiene in the $D_{\rm 4h}$ structure\cite{karadakov2008ground}. While cyclobutadiene
exhibits a first-order JT effect, in the closed-shell system 1AP the effect stems from second-order interactions. 
When including high-level and adiabatic corrections, our estimation of the STG of 1AP in the \cthreeh structure
explains the experimental value of 0.04 eV\cite{leupin1980low} better than the more negative STG of the \dthreeh structure. 
As the energy difference between the high and low-symmetry structures of 1AP is very small, future works can apply higher-level methods such as CCSDT or CCSDT(Q) for identifying the true minimum of 1AP.

5AP and 7AP with N atoms at electron-rich sites do not show symmetry lowering at the CCSD(T)-level and exhibit negative STG in ADC(2) calculations in line with the experimental  results\cite{ehrmaier2019singlet,wilson2024spectroscopic} and TBEs\cite{loos2023heptazine}. 
Three APs with N at the electron-deficient sites---2AP, 3AP, and 4AP---show stronger tendencies for symmetry lowering, even when modeled with a long-range corrected DFT method. While at the correct minima determined with CCSD(T), these molecules show positive STGs, 
popular methods such as MP2 and B3LYP prefer symmetric structures leading to negative STGs in subsequent excited state modeling,  suggesting a violation of Hund's rule. 
More rigorous studies should be conducted on the sensitivity of STGs to symmetry constraints of 
inverted-STG candidates not explored in this study. 
Our calculations show that 4AP exhibits significant pseudo-Jahn--Teller distortions, with the STG at the \cthreeh structure exceeding 0.3 eV, which is above the threshold for efficient 
thermally-activated-delayed-fluorescence
(TADF) emitters with rapid rISC\cite{samanta2017up,chen2018thermally,xie2020efficient}.
It is essential to ensure that novel properties capable of a paradigm shift in the theoretical understanding of molecules withstand the artifacts of quantum chemistry approximations.

\section{Supplementary Information}
i) Tables S1--S4 compare STGs from our calculations to results from past works; 
ii) Figure~S1 presents potential energy surface plots of S$_0$ calculated with DFT methods;
iii) Figure~S2 presents potential energy surface plots of S$_1$ and T$_1$ calculated with ADC(2); 
iv) Figure~S3 presents plots of HOMO and LUMO;
v) Figure~S4 presents NICS(1)$_{\rm iso}$ and NICS(1)$_{\rm zz}$ values;
vi) Frozen-core CCSD(T)/cc-pVTZ equilibrium coordinates and Z-matrix.

\section{Data Availability}
The data that support the findings of this study are
within the article and its supplementary material.

\section{Acknowledgments}
We acknowledge the support of the 
Department of Atomic Energy, Government
of India, under Project Identification No.~RTI~4007. 
All calculations have been performed using 
the Helios computer cluster, 
which is an integral part of the MolDis 
Big Data facility, 
TIFR Hyderabad \href{http://moldis.tifrh.res.in}{(http://moldis.tifrh.res.in)}.

\section{Author Declarations}

\subsection{Author contributions}
\noindent 
{\bf AM}: 
Conceptualization (main); 
Analysis (equal); 
Data collection (equal); 
Writing (main).
{\bf KJ}: 
Analysis (equal); 
Data collection (equal).
{\bf SD}: 
Analysis (equal); 
Data collection (equal).
{\bf RR}: Conceptualization (main); 
Analysis (equal); 
Data collection (equal); 
Funding acquisition; 
Project administration and supervision; 
Resources; 
Writing (main).

\subsection{Conflicts of Interest}
The authors have no conflicts of interest to disclose.

\section{References}
\bibliography{ref} 
\end{document}


\begin{table*}[!hptb]
\begin{threeparttable}
\centering
\caption{{\bf Results for 1AP:}
Energies of the
S$_1$ and T$_1$ states with respect to the S$_0$
ground state
along with the singlet-triplet gap, S$_1$-T$_1$,
are given in eV.  
Methods used for determining the
equilibrium geometries and their point groups are also stated. 
The barrier for automerization, $E^\ddagger$, determined with the method used for optimization
is stated in kJ/mol.
Results from other studies are included for comparison. 
 }
\small\addtolength{\tabcolsep}{1.2pt}
\begin{tabular}[t]{lll lll lll l}
\hline
\multicolumn{3}{l}{Geometry } & 
\multicolumn{1}{l}{} &
\multicolumn{4}{l}{Excited states} & 
\multicolumn{1}{l}{}   &
\multicolumn{1}{l}{Source} \\
\cline{1-3} \cline{5-8} 
\multicolumn{1}{l}{Method} & 
\multicolumn{1}{l}{Symmetry} &
\multicolumn{1}{l}{$E^\ddagger$}   &
\multicolumn{1}{l}{}   &
\multicolumn{1}{l}{Method}   &
\multicolumn{1}{l}{S$_1$} &
\multicolumn{1}{l}{T$_1$} &
\multicolumn{1}{l}{S$_1$-T$_1$} & 
\multicolumn{1}{l}{}   &
\multicolumn{1}{l}{} \\
\hline 
B3LYP & \dthreeh, \cthreeh &  0.0  & & ADC(2) & 1.004  & 1.146 & $-$0.143  & & this work\\ 
$\omega$B97X-D & \dthreeh, \cthreeh & 0.0  & & \multicolumn{1}{c}{\ditto}  & 1.027  & 1.166 & $-$0.139  & & \multicolumn{1}{c}{\ditto} \\ 
MP2 & \dthreeh, \cthreeh &  0.0 & & \multicolumn{1}{c}{\ditto}  & 0.997  & 1.138 & $-$0.141   & & \multicolumn{1}{c}{\ditto} \\ 
CCSD & \dthreeh &  3.6  & & \multicolumn{1}{c}{\ditto}  & 1.012  & 1.153 & $-$0.142  & & \multicolumn{1}{c}{\ditto} \\ 
 \multicolumn{1}{c}{\ditto}    & \cthreeh &  &  & \multicolumn{1}{c}{\ditto}  & 1.391  & 1.334 & $+$0.057   & & \multicolumn{1}{c}{\ditto}  \\
CCSD(T) & \dthreeh & 0.5  & &\multicolumn{1}{c}{\ditto}   & 0.988  & 1.134 & $-$0.146   & & \multicolumn{1}{c}{\ditto} \\ 
  \multicolumn{1}{c}{\ditto}     & \cthreeh &  &  &\multicolumn{1}{c}{\ditto}  &1.143  & 1.204 & $-$0.061   & & \multicolumn{1}{c}{\ditto} \\
B97-3C & &   &  & NEVPT2(12,12) & 1.244 & 1.288 & $-$0.044  & & \RRef{ricci2021singlet}\\
      & &   &  & RMS-CASPT2 & 0.89 & 0.97 & $-$0.08  & & \RRef{valverde2024computational}\\
MP2 & &   &  & CC2 & 1.047 &1.180  & $-$0.133  & & \RRef{tuvckova2022origin}\\
MP2,ADC(2) & &   &  & CC2(adiabatic) & 0.976 &1.117  & $-$0.141  & & \RRef{tuvckova2022origin}\\
MP2,ADC(2) & &   &  & CC2 (0-0) & 0.992 &1.068  & $-$0.076  & & \RRef{tuvckova2022origin}\\
$\omega$B97X-D& &   &  & EOM-CCSD &  &   & $-$0.072  & & \RRef{garner2023double}\\
CCSD(T) & \dthreeh &   &  & ADC(2) & 1.001 & 1.138 & $-$0.137  & & \RRef{loos2023heptazine}\\
CCSD(T)& \dthreeh &   &  & ADC(3)& 0.81 & 0.87 & $-$0.06  & & \RRef{loos2023heptazine}\\
CCSD(T)& \dthreeh &   &  & TBE & 0.979 & 1.110 & $-$0.131  & & \RRef{loos2023heptazine}\\
exp. & &   &  &exp. & 0.97 & 0.93 & +0.04  & & \RRef{leupin1980low}$^a$\\
\hline
\end{tabular}
 \begin{tablenotes}
  \item \footnotesize{$^a$ Using S$_1$=0.78 $\mu$m$^{-1}$ and T$_1$=0.75 $\mu$m$^{-1}$ 
  from \RRef{leupin1980low} multiplied by 1.2398 eV/$\mu$m$^{-1}$.}
  \end{tablenotes}
\end{threeparttable}
\label{tab_table1}
\end{table*}%

\begin{table*}[!hptb]
\begin{threeparttable}
\centering
\caption{{\bf Results for 5AP:}
Energies of the
S$_1$ and T$_1$ states with respect to the S$_0$
ground state
along with the singlet-triplet gap, S$_1$-T$_1$,
are given in eV.  
Methods used for determining the
equilibrium geometries and their point groups are also stated. 
The barrier for automerization, $E^\ddagger$, determined with the method used for optimization
is stated in kJ/mol.
Results from other studies are included for comparison. 
 }
\small\addtolength{\tabcolsep}{1.2pt}
\begin{tabular}[t]{lll lll lll l}
\hline
\multicolumn{3}{l}{Geometry } & 
\multicolumn{1}{l}{} &
\multicolumn{4}{l}{Excited states} & 
\multicolumn{1}{l}{}   &
\multicolumn{1}{l}{Source} \\
\cline{1-3} \cline{5-8} 
\multicolumn{1}{l}{Method} & 
\multicolumn{1}{l}{Symmetry} &
\multicolumn{1}{l}{$E^\ddagger$}   &
\multicolumn{1}{l}{}   &
\multicolumn{1}{l}{Method}   &
\multicolumn{1}{l}{S$_1$} &
\multicolumn{1}{l}{T$_1$} &
\multicolumn{1}{l}{S$_1$-T$_1$} & 
\multicolumn{1}{l}{}   &
\multicolumn{1}{l}{} \\
\hline 
B3LYP& \ctwov, \cs &  0.0  & & ADC(2) & 2.129  & 2.275 & $-$0.146   & & this work\\ 
$\omega$B97X-D & \ctwov, \cs & 0.0  & & \multicolumn{1}{c}{\ditto}  & 2.167  & 2.307 & $-$0.140  & & \multicolumn{1}{c}{\ditto}  \\ 
MP2 & \ctwov &   & & \multicolumn{1}{c}{\ditto}  & 2.143  & 2.288 & $-$0.145   & & \multicolumn{1}{c}{\ditto}  \\ 
CCSD & \ctwov &   & & \multicolumn{1}{c}{\ditto}  & 2.161  & 2.303 & $-$0.142   & & \multicolumn{1}{c}{\ditto}  \\ 
 CCSD(T) & \ctwov &   & & \multicolumn{1}{c}{\ditto}  & 2.123  & 2.273 & $-$0.150   & & \multicolumn{1}{c}{\ditto} \\ 
 B97-3C & &   &  & NEVPT2(8,8) &2.26  & 2.35 & $-$0.089  & & \RRef{ricci2021singlet}\\
B97-3C & &   &  & NEVPT2(10,10) &  &  & $+$0.034  & & \RRef{ricci2021singlet}\\
B97-3C & &   &  & NEVPT2(12,12) &  &  & $+$0.036  & & \RRef{ricci2021singlet}\\
      & &   &  & RMS-CASPT2 & 2.00 & 2.08 & $-$0.08  & & \RRef{valverde2024computational}\\
 MP2 & &   &  & CC2 & 2.231 &2.365  & $-$0.134  & & \RRef{tuvckova2022origin}\\
 MP2, ADC(2) & &   &  & CC2 (adiabatic) &2.066 &2.197  & $-$0.131  & & \RRef{tuvckova2022origin}\\
 MP2, ADC(2) & &   &  & CC2 (0-0) & 1.971 &2.056  & $-$0.085  & & \RRef{tuvckova2022origin}\\
 $\omega$B97X-D & &   &  & EOM-CCSD & 2.359 &2.397  & $-$0.08  & & \RRef{terence2023symmetry}\\
CCSD(T) & \ctwov &   &  & ADC(2) & 2.159 & 2.298 & $-$0.139  & & \RRef{loos2023heptazine}\\
CCSD(T)& \ctwov &   &  & TBE & 2.177 & 2.296 & $-$0.119  & & \RRef{loos2023heptazine}\\
 exp. & &   &  &exp. & 1.957 & 2.003 & $-$0.047  & & \RRef{wilson2024spectroscopic}\\
\hline
\end{tabular}
\end{threeparttable}
\label{tab_table1}
\end{table*}%

\begin{table*}[!hptb]
\begin{threeparttable}
\centering
\caption{{\bf Results for 7AP:}
Energies of the
S$_1$ and T$_1$ states with respect to the S$_0$
ground state
along with the singlet-triplet gap, S$_1$-T$_1$,
are given in eV.  
Methods used for determining the
equilibrium geometries and their point groups are also stated. 
The barrier for automerization, $E^\ddagger$, determined with the method used for optimization
is stated in kJ/mol.
Results from other studies are included for comparison. 
 }
\small\addtolength{\tabcolsep}{1.2pt}
\begin{tabular}[t]{lll lll lll l}
\hline
\multicolumn{3}{l}{Geometry } & 
\multicolumn{1}{l}{} &
\multicolumn{4}{l}{Excited states} & 
\multicolumn{1}{l}{}   &
\multicolumn{1}{l}{Source} \\
\cline{1-3} \cline{5-8} 
\multicolumn{1}{l}{Method} & 
\multicolumn{1}{l}{Symmetry} &
\multicolumn{1}{l}{$E^\ddagger$}   &
\multicolumn{1}{l}{}   &
\multicolumn{1}{l}{Method}   &
\multicolumn{1}{l}{S$_1$} &
\multicolumn{1}{l}{T$_1$} &
\multicolumn{1}{l}{S$_1$-T$_1$} & 
\multicolumn{1}{l}{}   &
\multicolumn{1}{l}{} \\
\hline 
B3LYP & \dthreeh,  \cthreeh &  0.0  & & ADC(2) & 2.643  & 2.892 & $-$0.249   & & this work\\ 
$\omega$B97X-D & \dthreeh,  \cthreeh & 0.0  & & \multicolumn{1}{c}{\ditto}  & 2.681  & 2.925 & $-$0.244   & &  \multicolumn{1}{c}{\ditto} \\ 
MP2 & \dthreeh &   & & \multicolumn{1}{c}{\ditto}  & 2.649  & 2.900 & $-$0.251   & & \multicolumn{1}{c}{\ditto}  \\ 
CCSD & \dthreeh &   & & \multicolumn{1}{c}{\ditto}  & 2.670  & 2.918 & $-$0.249   & & \multicolumn{1}{c}{\ditto}  \\ 
CCSD(T) & \dthreeh &   & & \multicolumn{1}{c}{\ditto}  & 2.626  & 2.882 & $-$0.256   & & \multicolumn{1}{c}{\ditto}  \\ 
B97-3C & &   &  & NEVPT2(12,12) & 3.259 & 3.398& $-$0.139  & & \RRef{ricci2021singlet}\\
  & &   &  & RMS-CASPT2 & 2.50 & 2.70 & $-$0.20  & & \RRef{valverde2024computational}\\
MP2 & &   &  & CC2 & 2.756 &2.998  & $-$0.242  & & \RRef{tuvckova2022origin}\\
MP2, ADC(2) & &   &  & CC2 (adiabatic) & 2.640 &2.894  & $-$0.254 & & \RRef{tuvckova2022origin}\\
MP2, ADC(2) & &   &  & CC2 (0-0) & 2.512 &2.618  & $-$0.106  & & \RRef{tuvckova2022origin}\\
 $\omega$B97X-D & &   &  & EOM-CCSD&  &   & $-$0.144  & & \RRef{garner2023double}\\
 CCSD(T) & \dthreeh &   &  & ADC(2) & 2.675 & 2.927 & $-$0.246  & & \RRef{loos2023heptazine}\\
 CCSD(T)& \dthreeh &   &  & ADC(3)& 2.81 & 2.88 & $-$0.07  & & \RRef{loos2023heptazine}\\
CCSD(T) & \dthreeh &   &  & TBE & 2.717 & 2.936 & $-$0.219  & & \RRef{loos2023heptazine}\\
 exp. & &   &  &exp. & &  & $<$0  & & \RRef{ehrmaier2019singlet}\\
\hline
\end{tabular}
\end{threeparttable}
\label{tab_table1}
\end{table*}%

\begin{table*}[!hptb]
\begin{threeparttable}
\centering
\caption{{\bf Results for 2AP, 3AP, and 4AP:}
Energies of the
S$_1$ and T$_1$ states with respect to the S$_0$
ground state
along with the singlet-triplet gap, S$_1$-T$_1$,
are given in eV.  
Methods used for determining the
equilibrium geometries and their point groups are also stated. 
The barrier for automerization, $E^\ddagger$, determined with the method used for optimization
is stated in kJ/mol.
Results from other studies are included for comparison.
 }
\small
\addtolength{\tabcolsep}{1.2pt}
\begin{tabular}[t]{lll lll lll l}
\hline
\multicolumn{3}{l}{Geometry } & 
\multicolumn{1}{l}{} &
\multicolumn{4}{l}{Excited states} & 
\multicolumn{1}{l}{}   &
\multicolumn{1}{l}{Source} \\
\cline{1-3} \cline{5-8} 
\multicolumn{1}{l}{Method} & 
\multicolumn{1}{l}{Symmetry} &
\multicolumn{1}{l}{$E^\ddagger$}   &
\multicolumn{1}{l}{}   &
\multicolumn{1}{l}{Method}   &
\multicolumn{1}{l}{S$_1$} &
\multicolumn{1}{l}{T$_1$} &
\multicolumn{1}{l}{S$_1$-T$_1$} & 
\multicolumn{1}{l}{}   &
\multicolumn{1}{l}{} \\
\hline 
\multicolumn{10}{l}{\bf Results for 2AP}\\
 B3LYP & \ctwov, \cs &  0.0  & & ADC(2) & 0.872  & 0.975 & $-$0.103  & & this work\\ 
 $\omega$B97X-D & \ctwov ($B_2$, 340$i$ cm$^{-1}$) & 0.1  & & \multicolumn{1}{c}{\ditto} & 0.894  & 0.993 & $-$0.099   & & \multicolumn{1}{c}{\ditto}\\ 
 \multicolumn{1}{c}{\ditto} & \cs &   &  & \multicolumn{1}{c}{\ditto} & 1.007  & 1.041 & $-$0.034   & & \multicolumn{1}{c}{\ditto}\\
 MP2 & \ctwov &  0.3 & & \multicolumn{1}{c}{\ditto} & 0.861  & 0.960 & $-$0.099   & & \multicolumn{1}{c}{\ditto}\\ 
 \multicolumn{1}{c}{\ditto} & \cs &   &  & \multicolumn{1}{c}{\ditto} & 0.851  & 0.948 & $-$0.097   & & \multicolumn{1}{c}{\ditto}\\
   CCSD & \ctwov & 9.6   & & \multicolumn{1}{c}{\ditto} & 0.885  & 0.988 & $-$0.102  & & \multicolumn{1}{c}{\ditto}\\ 
    \multicolumn{1}{c}{\ditto} & \cs &  &  & \multicolumn{1}{c}{\ditto} & 1.481  & 1.269 & $+$0.212 && \multicolumn{1}{c}{\ditto}\\   
 CCSD(T) & \ctwov & 2.5  & & \multicolumn{1}{c}{\ditto} & 0.856  & 0.962 & $-$0.106   & & \multicolumn{1}{c}{\ditto}\\ 
 \multicolumn{1}{c}{\ditto} & \cs &   &  & \multicolumn{1}{c}{\ditto}& 1.269  & 1.145 & $+$0.125   & & \multicolumn{1}{c}{\ditto}\\
B97-3C & &   &  & NEVPT2(12,12) &  &  & $-$0.120  & & \RRef{ricci2021singlet}\\
  & &   &  & RMS-CASPT2 & 0.89 & 0.97 & $-$0.08  & & \RRef{valverde2024computational}\\
 CCSD(T) & \ctwov &   &  & ADC(2) & 1.001 & 1.138 & $-$0.137  & & \RRef{loos2023heptazine}\\
 CCSD(T) & \ctwov &   &  & TBE & 0.833      & 0.904      & $-$0.071   & & \RRef{loos2023heptazine}\\
\hline 
\multicolumn{10}{l}{\bf Results for 3AP}\\
B3LYP & \ctwov,\cs &  0.0  & & ADC(2)& 0.743  & 0.827 & $-$0.084   & & this work\\ 
 $\omega$B97X-D & \ctwov ($B_2$, 773$i$ cm$^{-1}$) &  1.9 & & \multicolumn{1}{c}{\ditto} & 0.765  & 0.845 & $-$0.079  & & \multicolumn{1}{c}{\ditto}\\ 
 \multicolumn{1}{c}{\ditto} & \cs &  &  & \multicolumn{1}{c}{\ditto} & 1.238  & 1.049 & $+$0.189   & & \multicolumn{1}{c}{\ditto}\\
 MP2 & \ctwov,\cs &   & & \multicolumn{1}{c}{\ditto} & 0.711  & 0.790 & $-$0.079   & & \multicolumn{1}{c}{\ditto}\\ 
 CCSD & \ctwov &  16.8 & & \multicolumn{1}{c}{\ditto} & 0.745  & 0.826 & $-$0.081   & & \multicolumn{1}{c}{\ditto}\\ 
\multicolumn{1}{c}{\ditto} & \cs &   &  & \multicolumn{1}{c}{\ditto} & 1.555  & 1.225 & $+$0.330   & & \multicolumn{1}{c}{\ditto}\\
 CCSD(T) & \ctwov &  5.9  & & \multicolumn{1}{c}{\ditto} & 0.714  & 0.797 & $-$0.083   & & \multicolumn{1}{c}{\ditto}\\ 
  \multicolumn{1}{c}{\ditto} & \cs &  &  & \multicolumn{1}{c}{\ditto} & 1.367 & 1.093 & $+$0.274  & & \multicolumn{1}{c}{\ditto}\\
  CCSD(T) & \ctwov &   &  & ADC(2) & 2.159 & 2.298 & $-$0.139  & & \RRef{loos2023heptazine}\\
  CCSD(T) & \ctwov &   &  & TBE & 0.693       & 0.735      & $-$0.042 & & \RRef{loos2023heptazine}\\
\hline 
\multicolumn{10}{l}{\bf Results for 4AP}\\
B3LYP & \dthreeh ($A_2^\prime$, 494$i$ cm$^{-1}$)) &  0.2  & & ADC(2) & 0.614  & 0.689 & $-$0.075   & & this work\\ 
 \multicolumn{1}{c}{\ditto} & \cthreeh &   &  & \multicolumn{1}{c}{\ditto} & 0.916  & 0.791 & $+$0.125   & & \multicolumn{1}{c}{\ditto}\\
 $\omega$B97X-D & \dthreeh ($A_2^\prime$, 1113$i$ cm$^{-1}$) & 5.5  & & \multicolumn{1}{c}{\ditto} & 0.636  & 0.707 & $-$0.071   & & \multicolumn{1}{c}{\ditto}\\ 
 \multicolumn{1}{c}{\ditto} & \cthreeh &   &  & \multicolumn{1}{c}{\ditto} & 1.377  & 1.033 & $+$0.344   & & \multicolumn{1}{c}{\ditto}\\
 MP2 & \dthreeh,\cthreeh &   & & \multicolumn{1}{c}{\ditto} & 0.573  & 0.648 & $-$0.074   & & \multicolumn{1}{c}{\ditto}\\ 
   CCSD & \dthreeh & 26.3  & & \multicolumn{1}{c}{\ditto} & 0.614  & 0.688 & $-$0.073   & & \multicolumn{1}{c}{\ditto}\\ 
   \multicolumn{1}{c}{\ditto} & \cthreeh &   &  & \multicolumn{1}{c}{\ditto} & 1.635  & 1.203 & $+$0.432   & & \multicolumn{1}{c}{\ditto}\\
 CCSD(T) & \dthreeh & 11.6  & & \multicolumn{1}{c}{\ditto} & 0.582  & 0.660 & $-$0.078   & & \multicolumn{1}{c}{\ditto}\\ 
 \multicolumn{1}{c}{\ditto} & \cthreeh &   &  & \multicolumn{1}{c}{\ditto} & 1.452  & 1.057 & $+$0.395   & & \multicolumn{1}{c}{\ditto}\\
 CCSD(T) & \dthreeh &   &  & ADC(2) & 2.675 & 2.927 & $-$0.246  & & \RRef{loos2023heptazine}\\
 CCSD(T)& \dthreeh &   &  & TBE &  0.554      &    0.583   & $-$0.029  & & \RRef{loos2023heptazine}\\
\hline
\end{tabular}
\end{threeparttable}
\label{tab_2}
\end{table*}%

\begin{figure*}[ht]
    \centering
    \includegraphics[width=\linewidth]{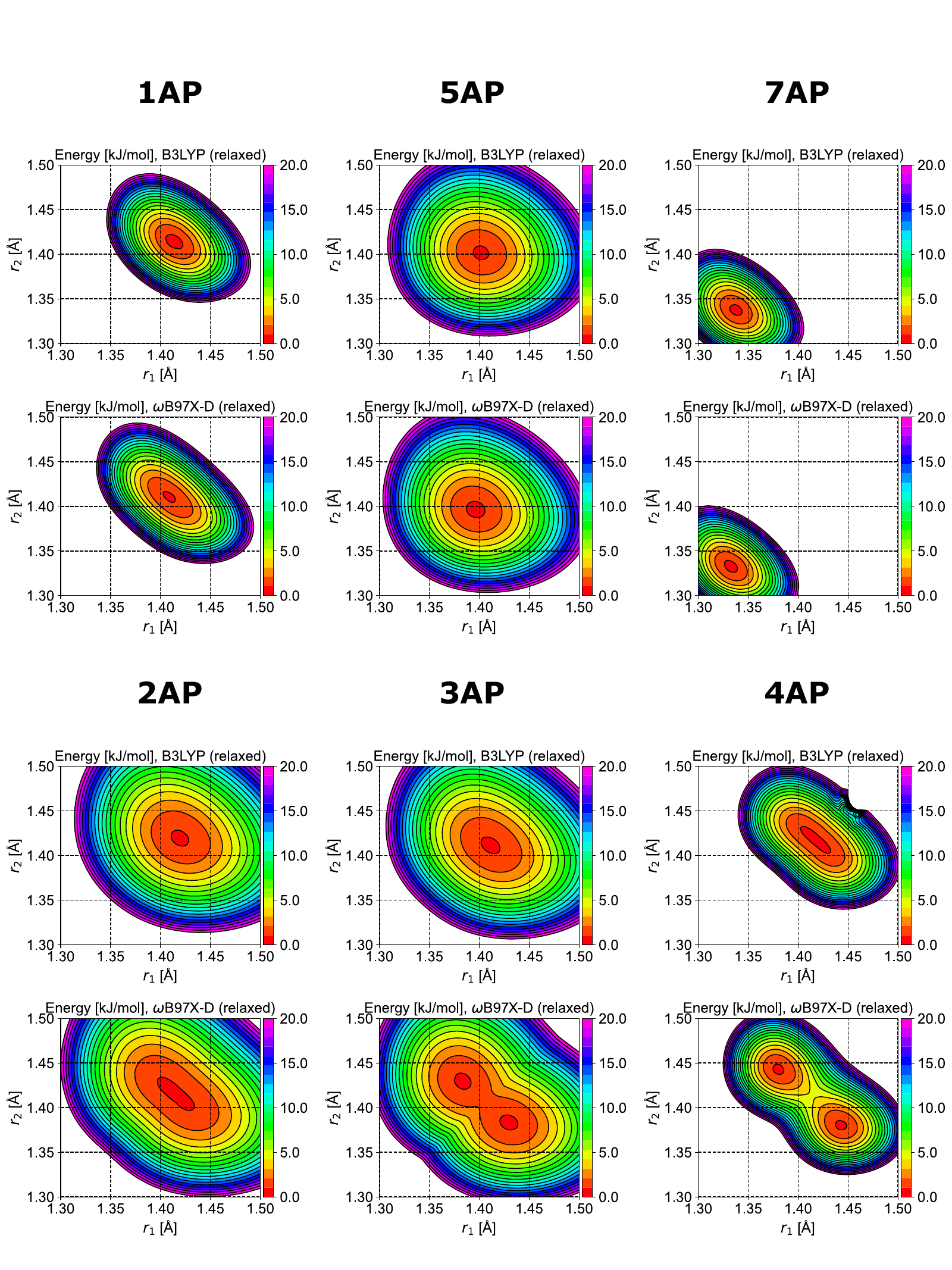}
    \caption{
Contour plots of two-dimensional potential energy surfaces 
of azaphenalenes calculated with the DFT methods: B3LYP and $\omega$B97X-D. See the main text for more details.}
\end{figure*}

\begin{figure*}[ht]
    \centering
    \includegraphics[width=\linewidth]{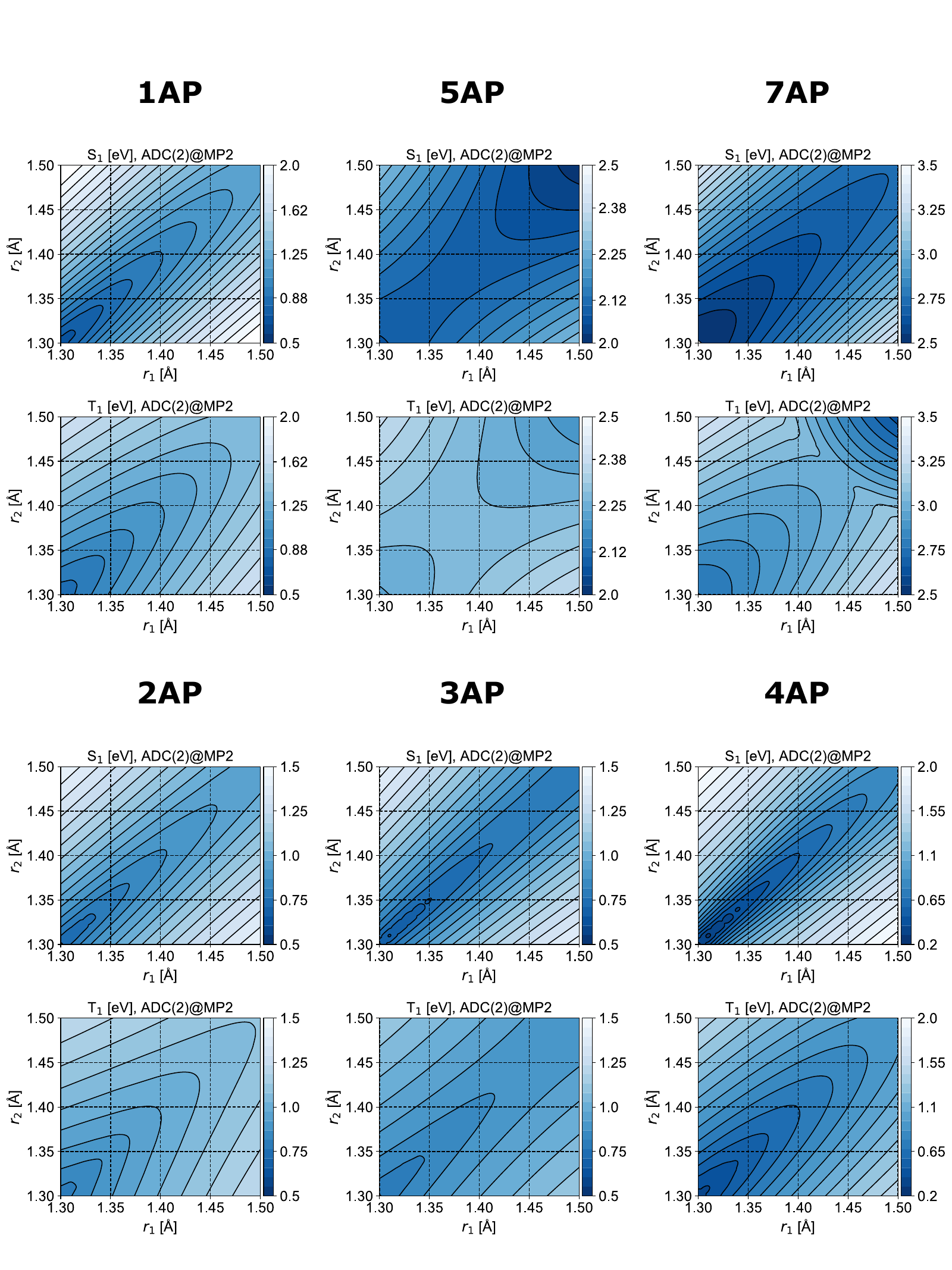}
    \caption{
Contour plots of S$_1$ and T$_1$ energies of azaphenalenes. See the main text for more details.}
\end{figure*}

\begin{figure*}[ht]
    \centering
    \includegraphics[width=\linewidth]{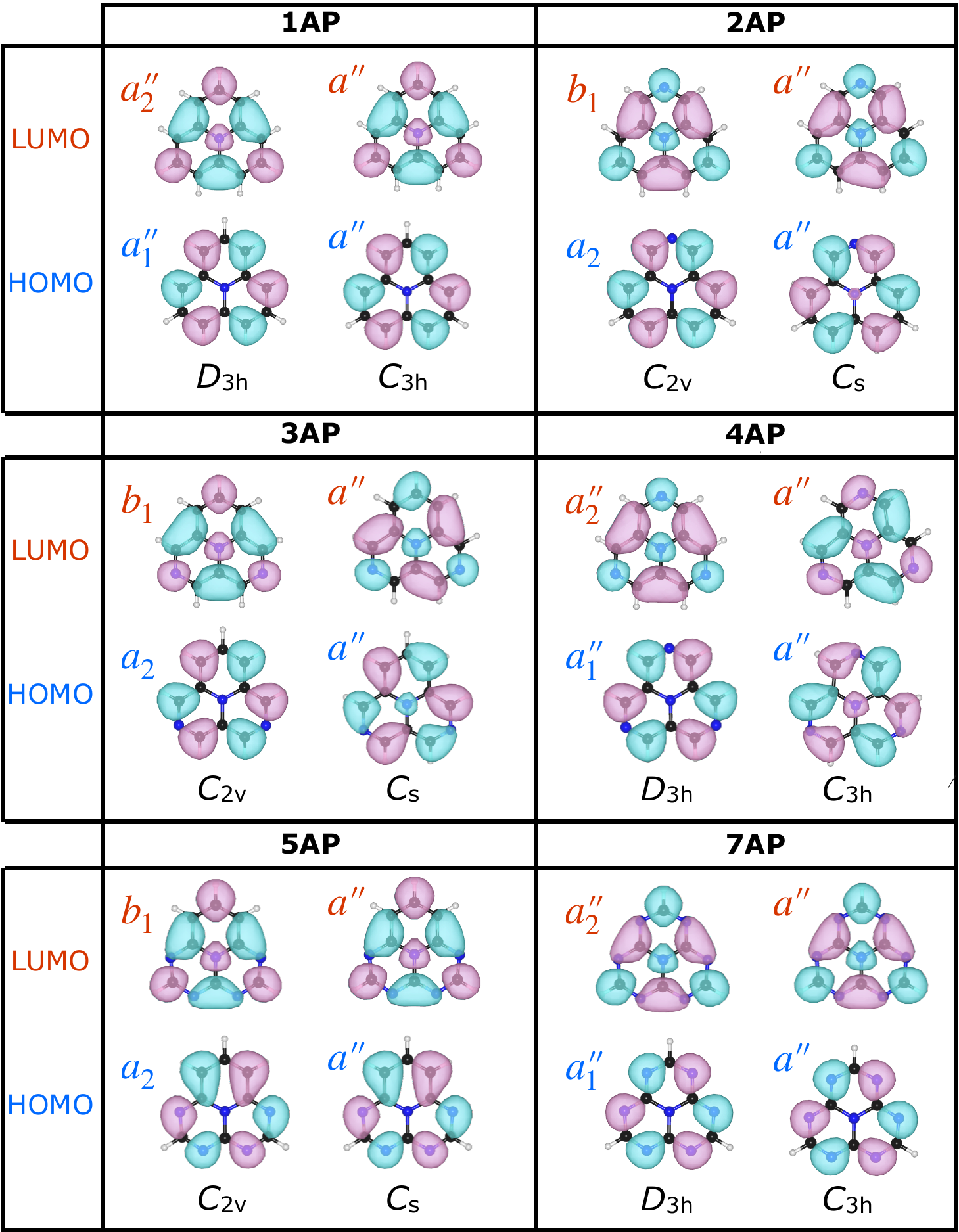}
    \caption{
Frontier MOs of azaphenalenes calculated with the
$\omega$B97X-D/cc-pVTZ method in the respective equilibrium geometries.
}
\end{figure*}

\begin{figure*}[ht]
    \centering
    \includegraphics[width=\linewidth]{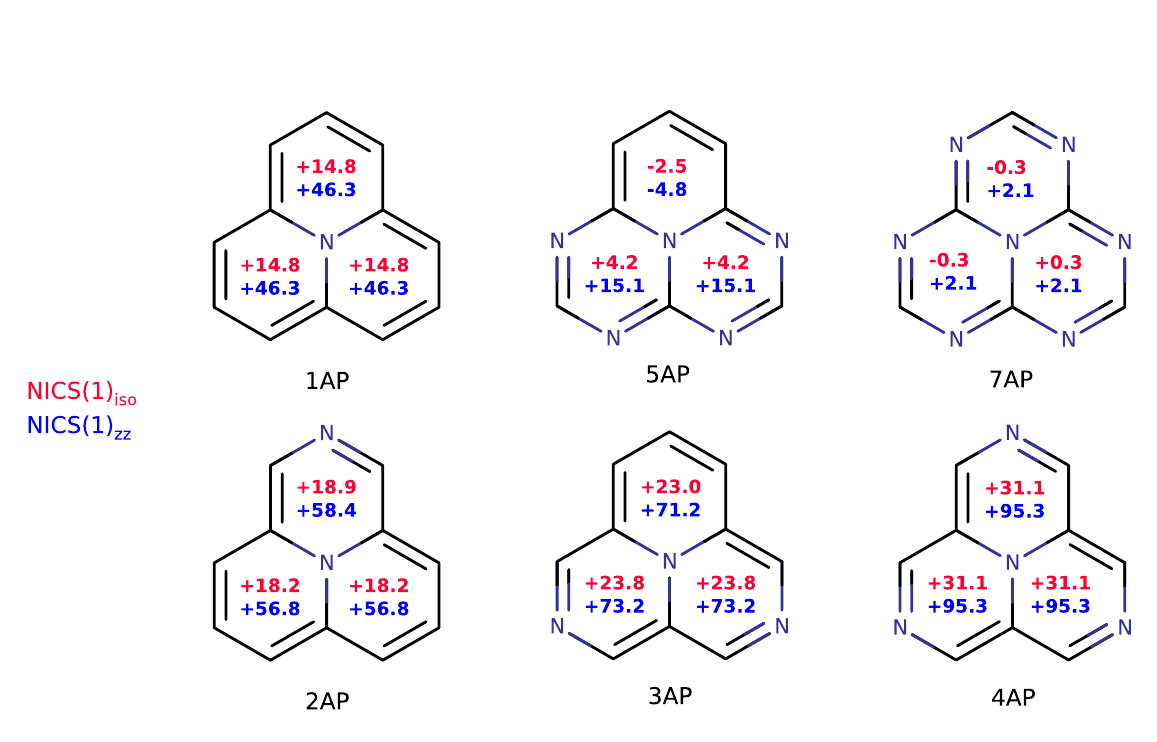}
    \caption{
Nucleus-independent chemical shifts (NICS) calculated with the
$\omega$B97X-D3/cc-pVTZ method.
}
\end{figure*}

\clearpage
\singlespacing
\small
{ 
\begin{verbatim}
-----------------------------------------------------------------
EQUILIBRIUM COORDINATES (ANGSTROEM), FROZEN-CORE CCSD(T)/cc-pVTZ 
MOLECULE: 1AP, POINT GROUP: D3H
-----------------------------------------------------------------
CARTESIAN COORDINATES
---------------------
   22
 CCSD(T)/CC-PVTZ  ENERGY=-516.45718151
 N         -0.0000000000        0.0000000000        0.0000000000
 C          1.4090613106        0.0000000000        0.0000000000
 C         -0.7045306553       -1.2202828904        0.0000000000
 C         -0.7045306553        1.2202828904        0.0000000000
 C          2.0972867575       -1.2322296003        0.0000000000
 C         -2.1157855159       -1.2001888109        0.0000000000
 C          0.0184987584        2.4324184112        0.0000000000
 C          1.4038057208       -2.4314628324        0.0000000000
 C         -2.8076114416        0.0000000000        0.0000000000
 C          1.4038057208        2.4314628324        0.0000000000
 C          0.0184987584       -2.4324184112        0.0000000000
 C         -2.1157855159        1.2001888109        0.0000000000
 C          2.0972867575        1.2322296003        0.0000000000
 H          3.1771396445       -1.1950402229        0.0000000000
 H         -2.6235050138       -2.1539635320        0.0000000000
 H         -0.5536346307        3.3490037549        0.0000000000
 H          1.9455458220       -3.3697842121        0.0000000000
 H         -3.8910916439        0.0000000000        0.0000000000
 H          1.9455458220        3.3697842121        0.0000000000
 H         -0.5536346307       -3.3490037549        0.0000000000
 H         -2.6235050138        2.1539635320        0.0000000000
 H          3.1771396445        1.1950402229        0.0000000000
-----------------------------------------------------------------
Z-MATRIX
--------
geometry={
N1
C1   ,  N1   ,  R1
C2   ,  N1   ,  R1  ,  C1  ,  120.0
C3   ,  N1   ,  R1  ,  C1  ,  120.0  ,  C2  ,  180.0
C4   ,  C1   ,  R2  ,  N1  ,  A1     ,  C3  ,  180.0
C5   ,  C2   ,  R2  ,  N1  ,  A1     ,  C1  ,  180.0
C6   ,  C3   ,  R2  ,  N1  ,  A1     ,  C2  ,  180.0
C7   ,  N1   ,  R3  ,  C1  ,  60.0   ,  C4  ,  0.0
C8   ,  N1   ,  R3  ,  C2  ,  60.0   ,  C5  ,  0.0
C9   ,  N1   ,  R3  ,  C3  ,  60.0   ,  C6  ,  0.0
C10  ,  C2   ,  R2  ,  N1  ,  A1     ,  C3  ,  180.0
C11  ,  C3   ,  R2  ,  N1  ,  A1     ,  C1  ,  180.0
C12  ,  C1   ,  R2  ,  N1  ,  A1     ,  C2  ,  180.0
H13  ,  C4   ,  R4  ,  C1  ,  A2     ,  N1  ,  180.0
H14  ,  C5   ,  R4  ,  C2  ,  A2     ,  N1  ,  180.0
H15  ,  C6   ,  R4  ,  C3  ,  A2     ,  N1  ,  180.0
H16  ,  N1   ,  R5  ,  C1  ,  60.0   ,  C4  ,  0.0
H17  ,  N1   ,  R5  ,  C2  ,  60.0   ,  C5  ,  0.0
H18  ,  N1   ,  R5  ,  C3  ,  60.0   ,  C6  ,  0.0
H19  ,  C10  ,  R4  ,  C2  ,  A2     ,  N1  ,  180.0
H20  ,  C11  ,  R4  ,  C3  ,  A2     ,  N1  ,  180.0
H21  ,  C12  ,  R4  ,  C1  ,  A2     ,  N1  ,  180.0}

R1=  1.40906131    ANG
R2=  1.41139791    ANG
R3=  2.80761144    ANG
R4=  1.08049309    ANG
R5=  3.89109164    ANG
A1=  119.18425212  DEGREE
A2=  117.21180516  DEGREE
-----------------------------------------------------------------

\end{verbatim}
}

\clearpage
\singlespacing
\small
{ 
\begin{verbatim}
-----------------------------------------------------------------
EQUILIBRIUM COORDINATES (ANGSTROEM), FROZEN-CORE CCSD(T)/cc-pVTZ 
MOLECULE: 1AP, POINT GROUP: C3H
-----------------------------------------------------------------
CARTESIAN COORDINATES
---------------------
   22
 CCSD(T)/CC-PVTZ  ENERGY=-516.45737101
 N         -0.0000000000       -0.0000000000        0.0000000000
 C          1.4102958789        0.0000000000        0.0000000000
 C         -0.7051479395       -1.2213520580        0.0000000000
 C         -0.7051479395        1.2213520580        0.0000000000
 C          2.0874818807       -1.2604285404        0.0000000000
 C         -2.1353040760       -1.1775980685        0.0000000000
 C          0.0478221953        2.4380266088        0.0000000000
 C          1.3925819281       -2.4404907466        0.0000000000
 C         -2.8098179483        0.0142340467        0.0000000000
 C          1.4172360202        2.4262566999        0.0000000000
 C         -0.0118165536       -2.4277621932        0.0000000000
 C         -2.0965954569        1.2241145322        0.0000000000
 C          2.1084120105        1.2036476610        0.0000000000
 H          3.1682050232       -1.2317091765        0.0000000000
 H         -2.6507939485       -2.1278914463        0.0000000000
 H         -0.5174110747        3.3596006228        0.0000000000
 H          1.9254527231       -3.3837087737        0.0000000000
 H         -3.8931041186        0.0243634149        0.0000000000
 H          1.9676513955        3.3593453588        0.0000000000
 H         -0.5905645004       -3.3398001629        0.0000000000
 H         -2.5970695345        2.1813439414        0.0000000000
 H          3.1876340349        1.1584562216        0.0000000000
-----------------------------------------------------------------
Z-MATRIX
--------
geometry={
N1
C1   ,  N1   ,  R1
C2   ,  N1   ,  R1   ,  C1  ,  120.0
C3   ,  N1   ,  R1   ,  C1  ,  120.0  ,  C2  ,  180.0
C4   ,  C1   ,  R21  ,  N1  ,  A11    ,  C3  ,  180.0
C5   ,  C2   ,  R21  ,  N1  ,  A11    ,  C1  ,  180.0
C6   ,  C3   ,  R21  ,  N1  ,  A11    ,  C2  ,  180.0
C7   ,  N1   ,  R3   ,  C1  ,  A3     ,  C4  ,  0.0
C8   ,  N1   ,  R3   ,  C2  ,  A3     ,  C5  ,  0.0
C9   ,  N1   ,  R3   ,  C3  ,  A3     ,  C6  ,  0.0
C10  ,  C2   ,  R22  ,  N1  ,  A12    ,  C3  ,  180.0
C11  ,  C3   ,  R22  ,  N1  ,  A12    ,  C1  ,  180.0
C12  ,  C1   ,  R22  ,  N1  ,  A12    ,  C2  ,  180.0
H13  ,  C4   ,  R41  ,  C1  ,  A21    ,  N1  ,  180.0
H14  ,  C5   ,  R41  ,  C2  ,  A21    ,  N1  ,  180.0
H15  ,  C6   ,  R41  ,  C3  ,  A21    ,  N1  ,  180.0
H16  ,  N1   ,  R5   ,  C1  ,  A4     ,  C4  ,  0.0
H17  ,  N1   ,  R5   ,  C2  ,  A4     ,  C5  ,  0.0
H18  ,  N1   ,  R5   ,  C3  ,  A4     ,  C6  ,  0.0
H19  ,  C10  ,  R42  ,  C2  ,  A22    ,  N1  ,  180.0
H20  ,  C11  ,  R42  ,  C3  ,  A22    ,  N1  ,  180.0
H21  ,  C12  ,  R42  ,  C1  ,  A22    ,  N1  ,  180.0}

R1=   1.41029588    ANG
R21=  1.43082528    ANG
R22=  1.39145026    ANG
R3=   2.80985400    ANG
R41=  1.08110467    ANG
R42=  1.08016779    ANG
R5=   3.89318035    ANG
A11=  118.24764782  DEGREE
A12=  120.11375054  DEGREE
A21=  116.72541599  DEGREE
A22=  117.71594276  DEGREE
A3=   60.29024792   DEGREE
A4=   60.35855774   DEGREE
-----------------------------------------------------------------
\end{verbatim}
}

\clearpage
\singlespacing
\small
{ 
\begin{verbatim}
-----------------------------------------------------------------
EQUILIBRIUM COORDINATES (ANGSTROEM), FROZEN-CORE CCSD(T)/cc-pVTZ 
MOLECULE: 5AP, POINT GROUP: C2V
-----------------------------------------------------------------
CARTESIAN COORDINATES
---------------------
   18
 CCSD(T)/CC-PVTZ  ENERGY=-580.61455705
 N          0.0000000000        0.0000000000       -0.0209337139
 C          0.0000000000        0.0000000000        1.3959162652
 C          0.0000000000        1.2239982977       -0.6919457558
 C          0.0000000000       -1.2239982977       -0.6919457558
 C          0.0000000000        1.2095528292       -2.0899593840
 C          0.0000000000       -1.2095528292       -2.0899593840
 C          0.0000000000        0.0000000000       -2.7748568006
 N          0.0000000000        2.3932182796       -0.0052415583
 N          0.0000000000       -2.3932182796       -0.0052415583
 C          0.0000000000        2.2752712648        1.3127808253
 C          0.0000000000       -2.2752712648        1.3127808253
 N          0.0000000000       -1.1592137265        2.0571477735
 N          0.0000000000        1.1592137265        2.0571477735
 H          0.0000000000        2.1667462052       -2.5888851619
 H          0.0000000000       -2.1667462052       -2.5888851619
 H          0.0000000000        0.0000000000       -3.8578325493
 H          0.0000000000        3.2048943207        1.8718610719
 H          0.0000000000       -3.2048943207        1.8718610719
-----------------------------------------------------------------
Z-MATRIX
--------
geometry={
Q1
N2   ,  Q1   ,  1.0
C3   ,  N2   ,  R1   ,Q1   ,  90.0
C4   ,  N2   ,  R2   ,C3   ,  A1    ,  Q1  ,  90.0
C5   ,  N2   ,  R2   ,C3   ,  A1    ,  Q1  ,  -90.0
C6   ,  C4   ,  R3   ,N2   ,  A2    ,  C3  ,  180.0
C7   ,  C5   ,  R3   ,N2   ,  A2    ,  C3  ,  180.0
C8   ,  N2   ,  R4   ,Q1   ,  90.0  ,  C3  ,  180.0
N9   ,  C4   ,  R5   ,C6   ,  A3    ,  C8  ,  180.0
N10  ,  C5   ,  R5   ,C7   ,  A3    ,  C8  ,  180.0
C11  ,  N9   ,  R6   ,C4   ,  A4    ,  C6  ,  180.0
C12  ,  N10  ,  R6   ,C5   ,  A4    ,  C7  ,  180.0
N13  ,  C3   ,  R7   ,N2   ,  A5    ,  Q1  ,  90.0
N14  ,  C3   ,  R7   ,N2   ,  A5    ,  Q1  ,  -90.0
H15  ,  C6   ,  R8   ,C4   ,  A6    ,  N2  ,  180.0
H16  ,  C7   ,  R8   ,C5   ,  A6    ,  N2  ,  180.0
H17  ,  N2   ,  R9   ,Q1   ,  90.0  ,  C3  ,  180.0
H18  ,  C11  ,  R10  ,N9   ,  A7    ,  C4  ,  180.0
H19  ,  C12  ,  R10  ,N10  ,  A7    ,  C5  ,  180.0}

R1=   1.41684998    ANG
R2=   1.39586138    ANG
R3=   1.39808826    ANG
R4=   2.75392309    ANG
R5=   1.35596387    ANG
R6=   1.32328927    ANG
R7=   1.33454246    ANG
R8=   1.07941933    ANG
R9=   3.83689884    ANG
R10=  1.08479019    ANG
A1=   118.73213541  DEGREE
A2=   118.14012765  DEGREE
A3=   121.01850011  DEGREE
A4=   115.31283627  DEGREE
A5=   119.70103198  DEGREE
A6=   116.93824030  DEGREE
A7=   115.90926862  DEGREE
 -----------------------------------------------------------------

\end{verbatim}
}

\clearpage
\singlespacing
\small
{ 
\begin{verbatim}
-----------------------------------------------------------------
EQUILIBRIUM COORDINATES (ANGSTROEM), FROZEN-CORE CCSD(T)/cc-pVTZ 
MOLECULE: 7AP, POINT GROUP: D3H
-----------------------------------------------------------------
CARTESIAN COORDINATES
---------------------
   16
 CCSD(T)/CC-PVTZ  ENERGY=-612.67978274
 N         -0.0000000000        0.0000000000        0.0000000000
 C          1.4025590960        0.0000000000        0.0000000000
 C         -0.7012795480       -1.2146518075        0.0000000000
 C         -0.7012795480        1.2146518075        0.0000000000
 N          2.0624112763       -1.1624036292        0.0000000000
 N         -2.0378767105       -1.2048987437        0.0000000000
 N         -0.0245345658        2.3673023730        0.0000000000
 C          1.3055871332       -2.2613432484        0.0000000000
 C         -2.6111742664        0.0000000000        0.0000000000
 C          1.3055871332        2.2613432484        0.0000000000
 N         -0.0245345658       -2.3673023730        0.0000000000
 N         -2.0378767105        1.2048987437        0.0000000000
 N          2.0624112763        1.1624036292        0.0000000000
 H          1.8477262446       -3.2003557341        0.0000000000
 H         -3.6954524892        0.0000000000        0.0000000000
 H          1.8477262446        3.2003557341        0.0000000000
-----------------------------------------------------------------
Z-MATRIX
--------
geometry={
N1
C1   ,  N1  ,  R1
C2   ,  N1  ,  R1  ,  C1  ,  120.0
C3   ,  N1  ,  R1  ,  C1  ,  120.0  ,  C2  ,  180.0
N4   ,  C1  ,  R2  ,  N1  ,  A1     ,  C3  ,  180.0
N5   ,  C2  ,  R2  ,  N1  ,  A1     ,  C1  ,  180.0
N6   ,  C3  ,  R2  ,  N1  ,  A1     ,  C2  ,  180.0
C7   ,  N1  ,  R3  ,  C1  ,  60.0   ,  N4  ,  0.0
C8   ,  N1  ,  R3  ,  C2  ,  60.0   ,  N5  ,  0.0
C9   ,  N1  ,  R3  ,  C3  ,  60.0   ,  N6  ,  0.0
N10  ,  C2  ,  R2  ,  N1  ,  A1     ,  C3  ,  180.0
N11  ,  C3  ,  R2  ,  N1  ,  A1     ,  C1  ,  180.0
N12  ,  C1  ,  R2  ,  N1  ,  A1     ,  C2  ,  180.0
H16  ,  N1  ,  R4  ,  C1  ,  60.0   ,  N4  ,  0.0
H17  ,  N1  ,  R4  ,  C2  ,  60.0   ,  N5  ,  0.0
H18  ,  N1  ,  R4  ,  C3  ,  60.0   ,  N6  ,  0.0}

R1=  1.40255910    ANG
R2=  1.33663275    ANG
R3=  2.61117427    ANG
R4=  3.69545249    ANG
A1=  119.58192379  DEGREE
-----------------------------------------------------------------
\end{verbatim}
}

\clearpage
\singlespacing
\small
{ 
\begin{verbatim}
-----------------------------------------------------------------
EQUILIBRIUM COORDINATES (ANGSTROEM), FROZEN-CORE CCSD(T)/cc-pVTZ 
MOLECULE: 2AP, POINT GROUP: C2V
-----------------------------------------------------------------
CARTESIAN COORDINATES
---------------------
   21
 CCSD(T)/CC-PVTZ  ENERGY=-532.47959209
 N          0.0000000000        0.0000000000        0.0248325005
 C          0.0000000000        0.0000000000        1.4290623782
 C          0.0000000000        1.2078318281       -0.6816722169
 C          0.0000000000       -1.2078318281       -0.6816722169
 C          0.0000000000        1.1361626851       -2.0998624488
 C          0.0000000000       -1.1361626851       -2.0998624488
 N          0.0000000000        0.0000000000       -2.7913610218
 C          0.0000000000        2.4312143725        0.0133915849
 C          0.0000000000       -2.4312143725        0.0133915849
 C          0.0000000000        2.4361423616        1.4014105437
 C          0.0000000000       -2.4361423616        1.4014105437
 C          0.0000000000       -1.2409740228        2.1053260238
 C          0.0000000000        1.2409740228        2.1053260238
 H          0.0000000000        2.0691835901       -2.6488864922
 H          0.0000000000       -2.0691835901       -2.6488864922
 H          0.0000000000        3.3405675958       -0.5697082153
 H          0.0000000000       -3.3405675958       -0.5697082153
 H          0.0000000000        3.3764935994        1.9393551417
 H          0.0000000000       -3.3764935994        1.9393551417
 H          0.0000000000       -1.2161284987        3.1855901804
 H          0.0000000000        1.2161284987        3.1855901804
-----------------------------------------------------------------
Z-MATRIX
--------
geometry={
Q1
N2   ,  Q1   ,  1.0
C3   ,  N2   ,  R1    ,  Q1   ,  90.0
C4   ,  N2   ,  R21   ,  C3   ,  A11   ,  Q1  ,  90.0
C5   ,  N2   ,  R21   ,  C3   ,  A11   ,  Q1  ,  -90.0
C6   ,  C4   ,  1.42  ,  N2   ,  A21   ,  C3  ,  180.0
C7   ,  C5   ,  1.42  ,  N2   ,  A21   ,  C3  ,  180.0
N8   ,  N2   ,  R4    ,  Q1   ,  90.0  ,  C3  ,  180.0
C9   ,  C4   ,  R51   ,  C6   ,  A31   ,  N8  ,  180.0
C10  ,  C5   ,  R51   ,  C7   ,  A31   ,  N8  ,  180.0
C11  ,  C9   ,  R61   ,  C4   ,  A41   ,  C6  ,  180.0
C12  ,  C10  ,  R61   ,  C5   ,  A41   ,  C7  ,  180.0
C13  ,  C3   ,  R71   ,  N2   ,  A51   ,  Q1  ,  90.0
C14  ,  C3   ,  R71   ,  N2   ,  A51   ,  Q1  ,  -90.0
H15  ,  C6   ,  R81   ,  C4   ,  A61   ,  N2  ,  180.0
H16  ,  C7   ,  R81   ,  C5   ,  A61   ,  N2  ,  180.0
H17  ,  C9   ,  R91   ,  C4   ,  A71   ,  N2  ,  180.0
H18  ,  C10  ,  R91   ,  C5   ,  A71   ,  N2  ,  180.0
H19  ,  C11  ,  R101  ,  C9   ,  A81   ,  C4  ,  180.0
H20  ,  C12  ,  R101  ,  C10  ,  A81   ,  C5  ,  180.0
H21  ,  C13  ,  R111  ,  C3   ,  A91   ,  N2  ,  180.0
H22  ,  C14  ,  R111  ,  C3   ,  A91   ,  N2  ,  180.0}

R1=    1.40422988    ANG
R21=   1.39928790    ANG
R4=    2.81619352    ANG
R51=   1.40704603    ANG
R61=   1.38802771    ANG
R71=   1.41327600    ANG
R81=   1.08256889    ANG
R91=   1.08024472    ANG
R101=  1.08334890    ANG
R111=  1.08054984    ANG
A11=   120.32491617  DEGREE
A21=   117.43189871  DEGREE
A31=   122.49605497  DEGREE
A41=   119.80645820  DEGREE
A51=   118.58799201  DEGREE
A61=   117.58111363  DEGREE
A71=   117.72796459  DEGREE
A81=   119.97590106  DEGREE
A91=   117.27045062  DEGREE
-----------------------------------------------------------------

\end{verbatim}
}

\clearpage
\singlespacing
\small
{ 
\begin{verbatim}
-----------------------------------------------------------------
EQUILIBRIUM COORDINATES (ANGSTROEM), FROZEN-CORE CCSD(T)/cc-pVTZ 
MOLECULE: 2AP, POINT GROUP: CS
-----------------------------------------------------------------
CARTESIAN COORDINATES
---------------------
   21
 CCSD(T)/CC-PVTZ  ENERGY=-532.48054565
 N          0.0000000000       -0.0045831784        0.0256744030
 C          0.0000000000       -0.0653222374        1.4328110801
 C          0.0000000000        1.2349703896       -0.6238470103
 C          0.0000000000       -1.1796125221       -0.7439882171
 C          0.0000000000        1.2001927918       -2.0721810380
 C          0.0000000000       -1.0755525634       -2.1230678573
 N          0.0000000000        0.1172406921       -2.7966089725
 C          0.0000000000        2.4209431626        0.0784696912
 C          0.0000000000       -2.4422851071       -0.0575750796
 C          0.0000000000        2.3859073808        1.4986888985
 C          0.0000000000       -2.4928065436        1.3058067208
 C          0.0000000000       -1.2937083640        2.0669099437
 C          0.0000000000        1.1868583471        2.1491593335
 H          0.0000000000        2.1601623484       -2.5787913104
 H          0.0000000000       -1.9796425986       -2.7154936531
 H          0.0000000000        3.3481710306       -0.4755702300
 H          0.0000000000       -3.3337207740       -0.6698562121
 H          0.0000000000        3.3094589364        2.0645423722
 H          0.0000000000       -3.4495452735        1.8138331151
 H          0.0000000000       -1.3101702046        3.1471666469
 H          0.0000000000        1.1246713808        3.2292659990
-----------------------------------------------------------------
Z-MATRIX
--------
geometry={
Q1
N2   ,  Q1   ,  1.0
C3   ,  N2   ,  R1    ,  Q1   ,  90.0
C4   ,  N2   ,  R21   ,  C3   ,  A11   ,  Q1  ,  90.0
C5   ,  N2   ,  R22   ,  C3   ,  A12   ,  Q1  ,  -90.0
C6   ,  C4   ,  R31   ,  N2   ,  A21   ,  C3  ,  180.0
C7   ,  C5   ,  R32   ,  N2   ,  A22   ,  C3  ,  180.0
N8   ,  N2   ,  R4    ,  Q1   ,  90.0  ,  C3  ,  180.0
C9   ,  C4   ,  R51   ,  C6   ,  A31   ,  N8  ,  180.0
C10  ,  C5   ,  R52   ,  C7   ,  A32   ,  N8  ,  180.0
C11  ,  C9   ,  R61   ,  C4   ,  A41   ,  C6  ,  180.0
C12  ,  C10  ,  R62   ,  C5   ,  A42   ,  C7  ,  180.0
C13  ,  C3   ,  R71   ,  N2   ,  A51   ,  Q1  ,  90.0
C14  ,  C3   ,  R72   ,  N2   ,  A52   ,  Q1  ,  -90.0
H15  ,  C6   ,  R81   ,  C4   ,  A61   ,  N2  ,  180.0
H16  ,  C7   ,  R82   ,  C5   ,  A62   ,  N2  ,  180.0
H17  ,  C9   ,  R91   ,  C4   ,  A71   ,  N2  ,  180.0
H18  ,  C10  ,  R92   ,  C5   ,  A72   ,  N2  ,  180.0
H19  ,  C11  ,  R101  ,  C9   ,  A81   ,  C4  ,  80.0
H20  ,  C12  ,  R102  ,  C10  ,  A82   ,  C5  ,  180.0
H21  ,  C13  ,  R111  ,  C3   ,  A91   ,  N2  ,  180.0
H22  ,  C14  ,  R112  ,  C3   ,  A92   ,  N2  ,  180.0}

R1=    1.40844697    ANG
R21=   1.39941813    ANG
R22=   1.40466171    ANG
R31=   1.44875151    ANG
R32=   1.38300005    ANG
R4=    2.82491142    ANG
R51=   1.37832513    ANG
R52=   1.43718651    ANG
R61=   1.42065130    ANG
R62=   1.36431754    ANG
R71=   1.38239424    ANG
R72=   1.44260564    ANG
R81=   1.08544715    ANG
R82=   1.08090107    ANG
R91=   1.08014432    ANG
R92=   1.08145538    ANG
R101=  1.08311478    ANG
R102=  1.08325427    ANG
R111=  1.08038213    ANG
R112=  1.08189539    ANG
A11=   120.12600781  DEGREE
A12=   120.75378845  DEGREE
A21=   116.27883973  DEGREE
A22=   118.91028829  DEGREE
A31=   122.00896152  DEGREE
A32=   122.84449320  DEGREE
A41=   119.22027278  DEGREE
A42=   120.65153460  DEGREE
A51=   119.77464009  DEGREE
A52=   117.30139889  DEGREE
A61=   116.44668394  DEGREE
A62=   118.92060510  DEGREE
A71=   118.50728259  DEGREE
A72=   116.98748256  DEGREE
A81=   120.08234665  DEGREE
A82=   120.09039442  DEGREE
A91=   118.17605456  DEGREE
A92=   116.47787965  DEGREE
-----------------------------------------------------------------
\end{verbatim}
}

\clearpage
\singlespacing
\small
{ 
\begin{verbatim}
-----------------------------------------------------------------
EQUILIBRIUM COORDINATES (ANGSTROEM), FROZEN-CORE CCSD(T)/cc-pVTZ 
MOLECULE: 3AP, POINT GROUP: C2V
-----------------------------------------------------------------
CARTESIAN COORDINATES
---------------------
   20
 CCSD(T)/CC-PVTZ  ENERGY=-548.50029036
 N          0.0000000000        0.0000000000       -0.0290948978
 C          0.0000000000        0.0000000000        1.3669984084
 C         -1.2168465284        0.0000000000       -0.7084797812
 C          1.2168465284        0.0000000000       -0.7084797812
 C         -1.2064367713        0.0000000000       -2.1185246596
 C          1.2064367713        0.0000000000       -2.1185246596
 C          0.0000000000        0.0000000000       -2.8062433640
 C         -2.4039234215        0.0000000000        0.0664921221
 C          2.4039234215        0.0000000000        0.0664921221
 N         -2.4222905439        0.0000000000        1.3968479847
 N          2.4222905439        0.0000000000        1.3968479847
 C          1.2485138341        0.0000000000        2.0294160803
 C         -1.2485138341        0.0000000000        2.0294160803
 H         -2.1584950794        0.0000000000       -2.6293185356
 H          2.1584950794        0.0000000000       -2.6293185356
 H          0.0000000000        0.0000000000       -3.8894859148
 H         -3.3500423056        0.0000000000       -0.4592347960
 H          3.3500423056        0.0000000000       -0.4592347960
 H          1.2501853399        0.0000000000        3.1116676060
 H         -1.2501853399        0.0000000000        3.1116676060
-----------------------------------------------------------------
Z-MATRIX
--------
geometry={
Q1
N2   ,  Q1   ,  1.
C3   ,  N2   ,  R1    ,  Q1  ,  90.
C4   ,  N2   ,  R21   ,  C3  ,  A11  ,  Q1  ,  90.
C5   ,  N2   ,  R21   ,  C3  ,  A11  ,  Q1  ,  -90.
C6   ,  C4   ,  R31   ,  N2  ,  A21  ,  C3  ,  180.
C7   ,  C5   ,  R31   ,  N2  ,  A21  ,  C3  ,  180.
C8   ,  N2   ,  R4    ,  Q1  ,  90.  ,  C3  ,  180.
C9   ,  C4   ,  R51   ,  C6  ,  A31  ,  C8  ,  180.
C10  ,  C5   ,  R51   ,  C7  ,  A31  ,  C8  ,  180.
N11  ,  C9   ,  R61   ,  C4  ,  A41  ,  C6  ,  180.
N12  ,  C10  ,  R61   ,  C5  ,  A41  ,  C7  ,  180.
C13  ,  C3   ,  R71   ,  N2  ,  A51  ,  Q1  ,  90.
C14  ,  C3   ,  R71   ,  N2  ,  A51  ,  Q1  ,  -90.
H15  ,  C6   ,  R81   ,  C4  ,  A61  ,  N2  ,  180.
H16  ,  C7   ,  R81   ,  C5  ,  A61  ,  N2  ,  180.
H17  ,  N2   ,  R9    ,  Q1  ,  90.  ,  C3  ,  180.
H18  ,  C9   ,  R101  ,  C4  ,  A71  ,  N2  ,  180.
H19  ,  C10  ,  R101  ,  C5  ,  A71  ,  N2  ,  180.
H20  ,  C13  ,  R111  ,  C3  ,  A81  ,  N2  ,  180.
H21  ,  C14  ,  R111  ,  C3  ,  A81  ,  N2  ,  180.}

R1=    1.39609331    ANG
R21=   1.39365681    ANG
R31=   1.41008330    ANG
R4=    2.77714847    ANG
R51=   1.41765052    ANG
R61=   1.33048265    ANG
R71=   1.41335911    ANG
R81=   1.08042834    ANG
R9=    3.86039102    ANG
R101=  1.08237227    ANG
R111=  1.08225282    ANG
A11=   119.17532185  DEGREE
A21=   118.75233935  DEGREE
A31=   123.56111295  DEGREE
A41=   123.92911561  DEGREE
A51=   117.94885423  DEGREE
A61=   117.79125752  DEGREE
A71=   117.80239208  DEGREE
A81=   118.03734581  DEGREE
-----------------------------------------------------------------
\end{verbatim}
}

\clearpage
\singlespacing
\small
{ 
\begin{verbatim}
-----------------------------------------------------------------
EQUILIBRIUM COORDINATES (ANGSTROEM), FROZEN-CORE CCSD(T)/cc-pVTZ 
MOLECULE: 3AP, POINT GROUP: CS
-----------------------------------------------------------------
CARTESIAN COORDINATES
---------------------
   20
 CCSD(T)/CC-PVTZ  ENERGY=-548.50254805
 N         -0.0301087996        0.0000000000       -0.0071919123
 C          1.3672653196        0.0000000000        0.0434018915
 C         -0.7451386686        0.0000000000        1.1985859497
 C         -0.6743830266        0.0000000000       -1.2451826933
 C         -2.1854701157        0.0000000000        1.1008388321
 C         -2.0478183588        0.0000000000       -1.3162538802
 C         -2.8111683861        0.0000000000       -0.1078839037
 C         -0.0667671869        0.0000000000        2.3955584988
 C          0.1946360907        0.0000000000       -2.4104311060
 N          1.3092378138        0.0000000000        2.4841624359
 N          1.4945797827        0.0000000000       -2.3753924311
 C          2.0868659377        0.0000000000       -1.1279470207
 C          1.9703973364        0.0000000000        1.3638484277
 H         -2.7395722113        0.0000000000        2.0300723981
 H         -2.5188281841        0.0000000000       -2.2885833186
 H         -3.8933635152        0.0000000000       -0.1470662294
 H         -0.6238735158        0.0000000000        3.3211015155
 H         -0.2953113948        0.0000000000       -3.3792420842
 H          3.1668701502        0.0000000000       -1.0931161855
 H          3.0551323907        0.0000000000        1.4020293484
-----------------------------------------------------------------
Z-MATRIX
--------
geometry={
Q1
N2   ,  Q1   ,  1.
C3   ,  N2   ,  R1    ,  Q1  ,  90.
C4   ,  N2   ,  R21   ,  C3  ,  A11  ,  Q1  ,  90.
C5   ,  N2   ,  R22   ,  C3  ,  A12  ,  Q1  ,  -90.
C6   ,  C4   ,  R31   ,  N2  ,  A21  ,  C3  ,  180.
C7   ,  C5   ,  R32   ,  N2  ,  A22  ,  C3  ,  180.
C8   ,  N2   ,  R4    ,  Q1  ,  90.  ,  C3  ,  180.
C9   ,  C4   ,  R51   ,  C6  ,  A31  ,  C8  ,  180.
C10  ,  C5   ,  R52   ,  C7  ,  A32  ,  C8  ,  180.
N11  ,  C9   ,  R61   ,  C4  ,  A41  ,  C6  ,  180.
N12  ,  C10  ,  R62   ,  C5  ,  A42  ,  C7  ,  180.
C13  ,  C3   ,  R71   ,  N2  ,  A51  ,  Q1  ,  90.
C14  ,  C3   ,  R72   ,  N2  ,  A52  ,  Q1  ,  -90.
H15  ,  C6   ,  R81   ,  C4  ,  A61  ,  N2  ,  180.
H16  ,  C7   ,  R82   ,  C5  ,  A62  ,  N2  ,  180.
H17  ,  N2   ,  R9    ,  Q1  ,  90.  ,  C3  ,  180.
H18  ,  C9   ,  R101  ,  C4  ,  A71  ,  N2  ,  180.
H19  ,  C10  ,  R102  ,  C5  ,  A72  ,  N2  ,  180.
H20  ,  C13  ,  R111  ,  C3  ,  A81  ,  N2  ,  180.
H21  ,  C14  ,  R112  ,  C3  ,  A82  ,  N2  ,  180.}

R1=    1.39828973    ANG
R21=   1.40184449    ANG
R22=   1.39560397    ANG
R31=   1.44364441    ANG
R32=   1.37527296    ANG
R4=    2.78288183    ANG
R51=   1.37583834    ANG
R52=   1.45361552    ANG
R61=   1.37885475    ANG
R62=   1.30041582    ANG
R71=   1.37473027    ANG
R72=   1.45167052    ANG
R81=   1.08189840    ANG
R82=   1.08040492    ANG
R9=    3.86578605    ANG
R101=  1.08027651    ANG
R102=  1.08565337    ANG
R111=  1.08056573    ANG
R112=  1.08540680    ANG
A11=   118.59450254  DEGREE
A12=   119.56690131  DEGREE
A21=   116.78568095  DEGREE
A22=   120.45558028  DEGREE
A31=   123.42432210  DEGREE
A32=   123.75260069  DEGREE
A41=   123.22624833  DEGREE
A42=   125.17086814  DEGREE
A51=   119.49029669  DEGREE
A52=   116.62276352  DEGREE
A61=   116.92524294  DEGREE
A62=   118.80843317  DEGREE
A71=   119.41333011  DEGREE
A72=   116.45850051  DEGREE
A81=   119.71667571  DEGREE
A82=   116.56508535  DEGREE
-----------------------------------------------------------------

\end{verbatim}
}

\clearpage
\singlespacing
\small
{ 
\begin{verbatim}
-----------------------------------------------------------------
EQUILIBRIUM COORDINATES (ANGSTROEM), FROZEN-CORE CCSD(T)/cc-pVTZ 
MOLECULE: 4AP, POINT GROUP: D3H
-----------------------------------------------------------------
CARTESIAN COORDINATES
---------------------
   19
 CCSD(T)/CC-PVTZ  ENERGY=-564.51919933
 N         -0.0000000000        0.0000000000        0.0000000000
 C          1.3895222367        0.0000000000        0.0000000000
 C         -0.6947611183       -1.2033615561        0.0000000000
 C         -0.6947611183        1.2033615561        0.0000000000
 C          2.0427788439       -1.2553814140        0.0000000000
 C         -2.1085816179       -1.1414076661        0.0000000000
 C          0.0658027740        2.3967890801        0.0000000000
 N          1.3980302792       -2.4214594740        0.0000000000
 N         -2.7960605583        0.0000000000        0.0000000000
 N          1.3980302792        2.4214594740        0.0000000000
 C          0.0658027740       -2.3967890801        0.0000000000
 C         -2.1085816179        1.1414076661        0.0000000000
 C          2.0427788439        1.2553814140        0.0000000000
 H          3.1249114130       -1.2687198584        0.0000000000
 H         -2.6611993341       -2.0718927391        0.0000000000
 H         -0.4637120789        3.3406125974        0.0000000000
 H         -0.4637120789       -3.3406125974        0.0000000000
 H         -2.6611993341        2.0718927391        0.0000000000
 H          3.1249114130        1.2687198584        0.0000000000
-----------------------------------------------------------------
Z-MATRIX
--------
geometry={
N1
C1  ,  N1  ,  R1
C2  ,  N1  ,  R1   ,  C1  ,  120.0
C3  ,  N1  ,  R1   ,  C1  ,  120.0  ,  C2  ,  180.0
C4  ,  C1  ,  R2   ,  N1  ,  A11    ,  C3  ,  180.0
C5  ,  C2  ,  R2   ,  N1  ,  A11    ,  C1  ,  180.0
C6  ,  C3  ,  R2   ,  N1  ,  A11    ,  C2  ,  180.0
N2  ,  N1  ,  R3   ,  C1  ,  60.0   ,  C4  ,  0.0
N3  ,  N1  ,  R3   ,  C2  ,  60.0   ,  C5  ,  0.0
N4  ,  N1  ,  R3   ,  C3  ,  60.0   ,  C6  ,  0.0
C7  ,  C2  ,  R2   ,  N1  ,  A11    ,  C3  ,  180.0
C8  ,  C3  ,  R2   ,  N1  ,  A11    ,  C1  ,  180.0
C9  ,  C1  ,  R2   ,  N1  ,  A11    ,  C2  ,  180.0
H1  ,  C4  ,  R41  ,  C1  ,  A21    ,  N1  ,  180.0
H2  ,  C5  ,  R41  ,  C2  ,  A21    ,  N1  ,  180.0
H3  ,  C6  ,  R41  ,  C3  ,  A21    ,  N1  ,  180.0
H4  ,  C7  ,  R41  ,  C2  ,  A21    ,  N1  ,  180.0
H5  ,  C8  ,  R41  ,  C3  ,  A21    ,  N1  ,  180.0
H6  ,  C9  ,  R41  ,  C1  ,  A21    ,  N1  ,  180.0}

R1=   1.38952224    ANG
R2=   1.41517726    ANG
R3=   2.79606056    ANG
R41=  1.08221477    ANG
A11=  117.49089295  DEGREE
A21=  118.19708911  DEGREE
-----------------------------------------------------------------
\end{verbatim}
}

\clearpage
\singlespacing
\small
{ 
\begin{verbatim}
-----------------------------------------------------------------
EQUILIBRIUM COORDINATES (ANGSTROEM), FROZEN-CORE CCSD(T)/cc-pVTZ 
MOLECULE: 4AP, POINT GROUP: C3H
-----------------------------------------------------------------
CARTESIAN COORDINATES
---------------------
   19
 CCSD(T)/CC-PVTZ  ENERGY=-564.52363202
 N         -0.0000000000        0.0000000000        0.0000000000
 C          1.3943168701        0.0000000000        0.0000000000
 C         -0.6971584351       -1.2075138304        0.0000000000
 C         -0.6971584351        1.2075138304        0.0000000000
 C          2.0196953303       -1.3136437071        0.0000000000
 C         -2.1474964870       -1.0922856104        0.0000000000
 C          0.1278011567        2.4059293175        0.0000000000
 N          1.3797047458       -2.4440919420        0.0000000000
 N         -2.8064980839        0.0271866114        0.0000000000
 N          1.4267933381        2.4169053306        0.0000000000
 C         -0.0029504215       -2.3888647060        0.0000000000
 C         -2.0673423109        1.1969874930        0.0000000000
 C          2.0702927324        1.1918772130        0.0000000000
 H          3.1047933775       -1.3373701978        0.0000000000
 H         -2.7105932543       -2.0201448396        0.0000000000
 H         -0.3942001232        3.3575150373        0.0000000000
 H         -0.5363847240       -3.3282397792        0.0000000000
 H         -2.6141478367        2.1286426868        0.0000000000
 H          3.1505325607        1.1995970924        0.0000000000
-----------------------------------------------------------------
Z-MATRIX
--------
geometry={
N1
C1  ,  N1  ,  R1
C2  ,  N1  ,  R1   ,  C1  ,  120.0
C3  ,  N1  ,  R1   ,  C1  ,  120.0  ,  C2  ,  180.0
C4  ,  C1  ,  R21  ,  N1  ,  A11    ,  C3  ,  180.0
C5  ,  C2  ,  R21  ,  N1  ,  A11    ,  C1  ,  180.0
C6  ,  C3  ,  R21  ,  N1  ,  A11    ,  C2  ,  180.0
N2  ,  N1  ,  R3   ,  C1  ,  A3     ,  C4  ,  0.0
N3  ,  N1  ,  R3   ,  C2  ,  A3     ,  C5  ,  0.0
N4  ,  N1  ,  R3   ,  C3  ,  A3     ,  C6  ,  0.0
C7  ,  C2  ,  R22  ,  N1  ,  A12    ,  C3  ,  180.0
C8  ,  C3  ,  R22  ,  N1  ,  A12    ,  C1  ,  180.0
C9  ,  C1  ,  R22  ,  N1  ,  A12    ,  C2  ,  180.0
H1  ,  C4  ,  R41  ,  C1  ,  A21    ,  N1  ,  180.0
H2  ,  C5  ,  R41  ,  C2  ,  A21    ,  N1  ,  180.0
H3  ,  C6  ,  R41  ,  C3  ,  A21    ,  N1  ,  180.0
H4  ,  C7  ,  R42  ,  C2  ,  A22    ,  N1  ,  180.0
H5  ,  C8  ,  R42  ,  C3  ,  A22    ,  N1  ,  180.0
H6  ,  C9  ,  R42  ,  C1  ,  A22    ,  N1  ,  180.0}

R1=   1.39431687    ANG
R21=  1.45490825    ANG
R22=  1.37022431    ANG
R3=   2.80662976    ANG
R41=  1.08535742    ANG
R42=  1.08026741    ANG
A11=  115.45743707  DEGREE
A12=  119.55983802  DEGREE
A21=  116.71005308  DEGREE
A22=  119.96929245  DEGREE
A3=   60.55500817   DEGREE
-----------------------------------------------------------------
\end{verbatim}
}

\onehalfspacing 
\clearpage 
\bibliography{ref}